\documentclass{article}

\usepackage{microtype}
\usepackage{graphicx}
\usepackage{booktabs} 
\usepackage{subfig}

\usepackage{hyperref}
\usepackage{multirow}

\usepackage[accepted]{icml2024}

\usepackage{amsmath}
\usepackage{amssymb}
\usepackage{mathtools}
\usepackage{amsthm}
\usepackage{makecell}

\usepackage[capitalize,noabbrev]{cleveref}

\theoremstyle{plain}

\theoremstyle{definition}

\theoremstyle{remark}

\usepackage{tikz}
\usetikzlibrary{arrows}
\usetikzlibrary{arrows.meta}

\tikzstyle{specialblock} = [draw, ultra thick, fill=blue!20, rectangle, 
    minimum height=3em, minimum width=4em]
\tikzstyle{block} = [draw, fill=lightgray, rectangle, 
    minimum height=3em, minimum width=4em]
\tikzstyle{sum} = [draw, fill=white, circle, node distance=1cm]
\tikzstyle{prod}   = [circle, minimum width=8pt, draw, inner sep=0pt, path picture={\draw (path picture bounding box.south east) -- (path picture bounding box.north west) (path picture bounding box.south west) -- (path picture bounding box.north east);}]
\tikzstyle{sumt}   = [circle, minimum width=8pt, draw, inner sep=0pt, path picture={\draw (path picture bounding box.east) -- (path picture bounding box.west) (path picture bounding box.south) -- (path picture bounding box.north);}]
\tikzstyle{input} = [coordinate]
\tikzstyle{output} = [coordinate]
\tikzstyle{pinstyle} = [pin edge={to-,thin,black}]
\tikzset{
tmp/.style  = {coordinate}, 
dot/.style = {circle, minimum size=#1,
              inner sep=0pt, outer sep=0pt},
dot/.default = 6pt 
}

\usepackage[textsize=tiny]{todonotes}

\icmltitlerunning{Listenable Maps for Audio Classifiers}

\begin{document}

\twocolumn[
\icmltitle{Listenable Maps for Audio Classifiers}




\begin{icmlauthorlist}
\icmlauthor{Francesco Paissan}{5,3}
\icmlauthor{Mirco Ravanelli}{2,3}
\icmlauthor{Cem Subakan}{4,2,3}
\end{icmlauthorlist}

\icmlaffiliation{5}{Fondazione Bruno Kessler}
\icmlaffiliation{2}{Concordia University}
\icmlaffiliation{3}{Mila-Québec AI Institute}
\icmlaffiliation{4}{Laval University}

\icmlcorrespondingauthor{Francesco Paissan}{fpaissan@fbk.eu}

\icmlkeywords{Machine Learning, ICML}

\vskip 0.3in
]



\printAffiliationsAndNotice{}  

\begin{abstract}


Despite the impressive performance of deep learning models across diverse tasks, their complexity poses challenges for interpretation. This challenge is particularly evident for audio signals, where conveying interpretations becomes inherently difficult.
To address this issue, we introduce Listenable Maps for Audio Classifiers (L-MAC),  a posthoc interpretation method that generates faithful and listenable interpretations. L-MAC utilizes a decoder on top of a pretrained classifier to generate binary masks that highlight relevant portions of the input audio. We train the decoder with a loss function that maximizes the confidence of the classifier decision on the masked-in portion of the audio while minimizing the probability of model output for the masked-out portion.
Quantitative evaluations on both in-domain and out-of-domain data demonstrate that L-MAC consistently produces more faithful interpretations than several gradient and masking-based methodologies. Furthermore, a user study confirms that, on average, users prefer the interpretations generated by the proposed technique.
\end{abstract}

\section{Introduction}
\label{sec:intro}

In recent years, deep learning models made significant strides in a variety of speech/audio applications, including sound event recognition, sound generation, speech recognition, speech separation, and many more \cite{ravanelli2021speechbrain}. 
An overwhelming majority of these models remain opaque concerning the interpretation of their predictions, as their large number of parameters, non-linearity, and high dimensionality make them "black-box" models \cite{molnar2022}.
Explainable Machine Learning is a research area that aims to render the models transparent concerning their decision-making mechanisms.
Posthoc interpretability methods \cite{smilkov2017smoothgrad, simonyan2014deep, l2i}, a sub-field within this domain, focus on generating interpretations for pre-trained machine learning models. These interpretations should ideally remain as faithful as possible to the pre-trained model while being easy to understand for humans.
Many existing posthoc interpretability methods are primarily designed for computer vision, where the task often involves classifying objects against a clean background. In these cases, the interpretation commonly takes the form of a saliency map, which highlights the regions of the image relevant to model predictions.
However, in the audio domain, achieving easily understandable interpretations poses a much greater challenge. State-of-the-art models for speech and audio processing typically operate on less interpretable inputs, such as mel-spectrograms, as compared to standard images. Consequently, generating saliency maps on these input features does not yield straightforward interpretations.
A potentially more promising but relatively underexplored alternative involves generating listenable interpretations, which offer a more natural and user-friendly way for humans to comprehend the model prediction.

This paper contributes to this emerging field by introducing a novel method called Listenable Maps for Audio Classifiers (L-MAC). L-MAC outputs listenable explanations for pretrained audio classifiers that utilize mel-spectrograms or any other feature as input.
Our approach employs a decoder that leverages information from the latent representations of pretrained classifiers to generate binary masks highlighting relevant audio segments. The decoder applies the mask not directly to the specific input features of the pretrained classifier but to the magnitude of Short-Time Fourier Transform (STFT) of the original input audio waveform.  By inheriting the phase from the original signal, we can perform the Inverse Short-Time Fourier Transform (ISTFT), allowing the generation of a listenable waveform as the outcome of the interpretation process. The decoder is trained with a loss function that maximizes the confidence of the classifier's decision on the masked-in portion of the audio while minimizing the probability of the model output for the masked-out portion. Our loss term explicitly guides the interpreter to produce explanations that closely follow the target source without sacrificing the faithfulness of the explanations.

\usetikzlibrary{arrows.meta}
\tikzstyle{dictsmall} = [draw, thick, fill=white!10, rectangle, 
    minimum height=1.0cm, minimum width=5cm] 
    \newcommand{\xshifts}{+4.7}
\begin{figure*}[t!]
    \centering
    \begin{tikzpicture}[auto, node distance=1.5cm,>=latex']
        \node [fill=none, xshift=0cm] (input) {$X$};

        \node [block, fill=white, right of=input, xshift=+.4cm, text width=1.3cm] (inpt1) {{InputTf}}; 
        \node [block, right of=inpt1, xshift=+.5cm] (cls) {Classifier}; 
        \node [right of=cls, xshift=.5cm] (h) {$h$}; 
        \node [below of=h, xshift=0cm, yshift=+.2cm] (ghst) {}; 
        \node [block, above of=h] (head) {OutputHead}; 
        \node [right of=head, xshift=.6cm] (yhat) {$\widehat y$};

        \node [block, fill=blue!10, right of=h, xshift=.5cm] (dec) {Decoder}; 
        \node [right of=dec, xshift=.3cm] (mask) {$M$}; 
        \node [prod, right of=mask, xshift=-.2cm] (prod) {}; 
        \node [block, fill=white, right of=prod, xshift=.15cm, text width=1.3cm] (cls2) { {{InputTf}}};
        \node [block, right of=cls2, xshift=.9cm, text width=1.7cm] (mloss) {\; Classifier\\ \; \; \; \;+\\ OutputHead};
        \node [above of=mloss, text width=1cm, yshift=.2cm] (mlossfr) {Masking \\\; Loss};
        \node [above of=prod, xshift=-.0cm, yshift=.2cm] (int) {\includegraphics[width=0.1\textwidth, trim=3cm 1.1cm 3.5cm 0cm, clip]{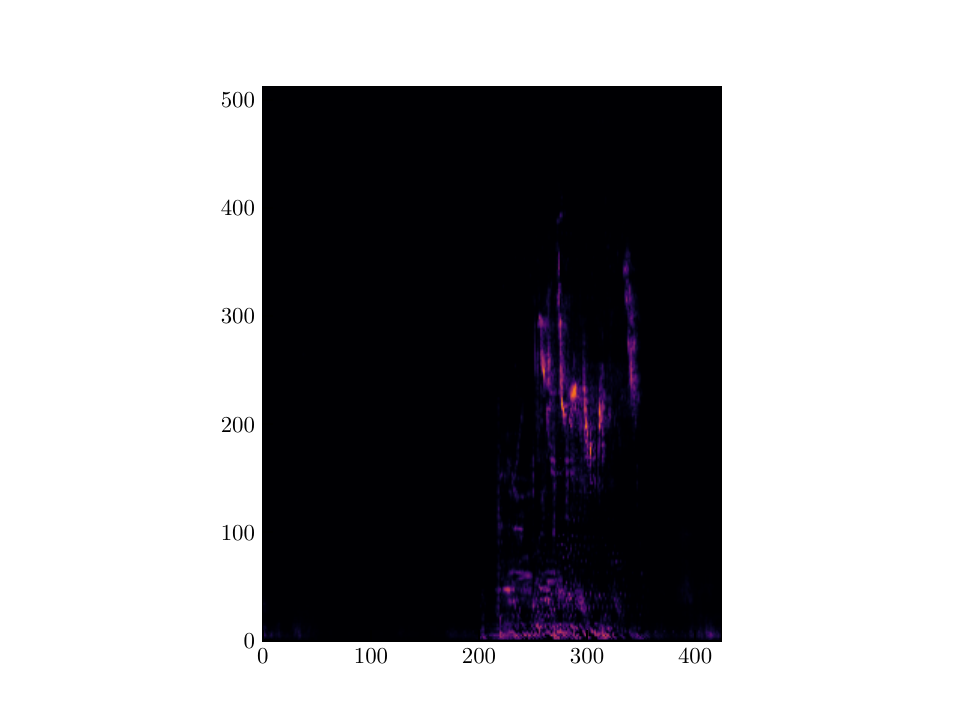}};
        \node [right of=int, text width=1.8cm, yshift=.3cm, xshift=1.2cm] (list) {\footnotesize{Listenable Interp.}};

        \node [above of=int, yshift=-.6cm] (ghsti) {}; 
        
        \draw [->, line width=1.2pt] (input) |- (ghst.center) [->] -| (prod);
        \draw [->, line width=1.5pt] (input) -- (inpt1);
        \draw [->, line width=1.5pt] (inpt1) -- (cls);
        \draw [->, line width=1.5pt] (cls) -- (h);
        \draw [->, line width=1.5pt] (h) -- (head);
        \draw [->, line width=1.5pt] (h) -- (dec);
        \draw [->, line width=1.5pt] (dec) -- (mask);
        \draw [->, line width=1.5pt] (mask) -- (prod);
        \draw [->, line width=1.5pt] (prod) -- (cls2);
        \draw [->, line width=1.5pt] (cls2) -- (mloss);
        \draw [->, line width=1.5pt] (mloss) -- (mlossfr);
        \draw [->, line width=1.5pt] (head) -- (yhat);
        \node [draw=none, fill=none, right of=input, yshift=.2cm, xshift=-1.7cm] (inppic)  {\includegraphics[width=0.1\textwidth, trim=3cm 1.1cm 3.5cm 0cm, clip]{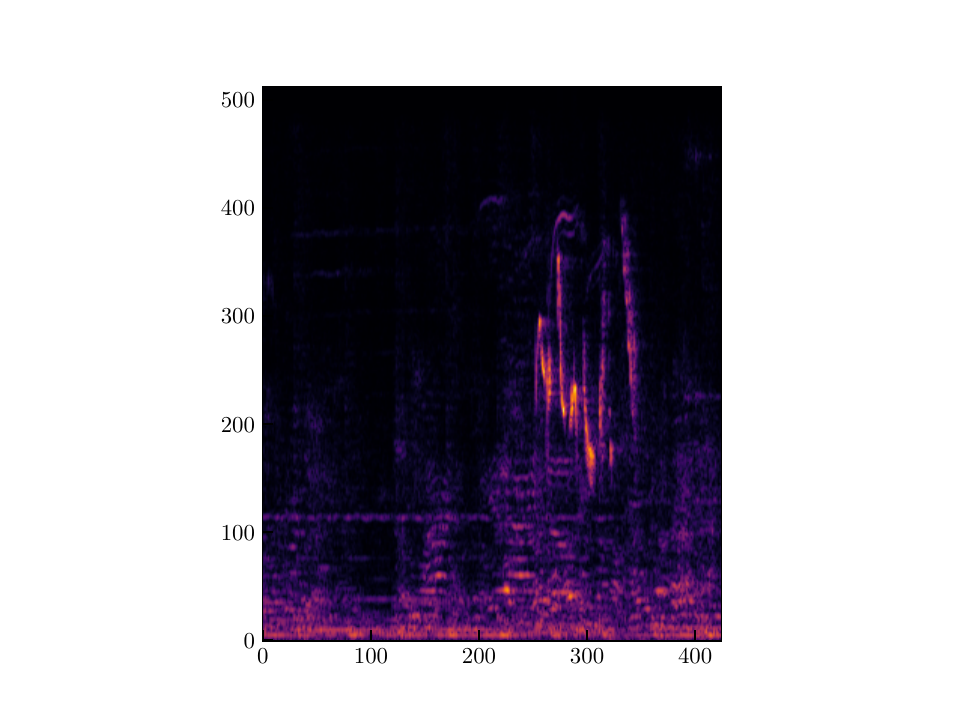} };
        \node[above of=inppic, yshift=-.4cm,xshift=.4cm] (X) {$X$}; 
        \draw [->, line width=1.5pt] (prod) -- (int.south);
        \draw [->, line width=1.5pt] (int) -- node [xshift=.5cm, yshift=.2cm] {\tiny{ISTFT}} (list);


    \end{tikzpicture}
    \caption{L-MAC Architecture. First, the linear spectrogram $X$ is computed from an audio waveform $x$. Then,  the audio features used by the pretrained classifier (e.g., FBANKs) are extracted (Input Tf). The classifier generates class predictions $\hat{y}$  and its latent representations $h$ are input to the decoder, which produces a binary mask $M$ for selecting specific portions of the original linear spectrogram $X$. The listenable interpretation is generated by applying the Inverse Short-Time Fourier Transform (ISTFT) on the masked spectrogram $X \odot M$ with the phase inherited from the original audio waveform. The masked loss used to train the decoder is computed based on the classifier predictions on the masked spectrogram and the predicted class $\hat{y}.$}
    \label{fig:arch}
    \vspace{-0.5cm}
\end{figure*}
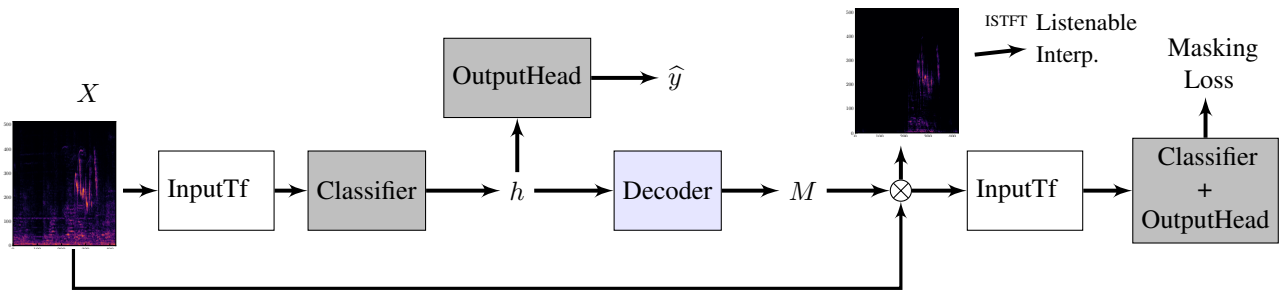

To summarize, our contributions are the following: 

\begin{itemize}
\item We propose a masking-based posthoc interpretation method for audio classifiers capable of providing listenable interpretations, even when the input audio is in the logmel domain.

\item Using various faithfulness metrics, we demonstrate that our proposed method outperforms several existing posthoc interpretation methods on both in-domain and out-of-domain data.

\item We conducted a user study, highlighting that users, on average, prefer the interpretation provided by L-MAC.

\item We illustrate that our proposed method allows fine-tuning during training to improve audio quality. We show that this does not result in substantial loss in faithfulness.

\end{itemize}


\newcommand{\cem}[1]{$\mathcal{CEM}:$ \textbf{#1}}
\newcommand{\cemcamera}[1]{{\color{black}{#1}}}

\subsection{Related Works}
In the literature, various methods have been proposed for generating saliency maps, which aim to highlight the portions of the input that significantly contribute to the model prediction. 
These methods can be broadly categorized into masking-based and gradient-based approaches. 

Gradient-based methods include techniques such as the standard saliency method, where a map is obtained by computing the gradient with respect to the input of the network, as described in \cite{simonyan2014deep}. Other methods include integrated gradients \cite{sundararajan2017axiomatic}, guided back-prop \cite{springenberg2015striving}, GradCAM \cite{Selvaraju_2019}, and SmoothGrad \cite{smilkov2017smoothgrad}. 
In \cite{adebayo2020sanity}, it has been shown that gradient-based approaches may not accurately capture the behavior of the classifier. Even under network weight and label randomization, these approaches essentially yield a behavior comparable to edge detection.

An emerging alternative is represented by masking-based approaches, which are the focus of this paper. Masking-based approaches involve estimating a mask (typically binary) to select a portion of the image that maximally contributes to the classifier decision. There exists some methods which directly learn a mask \cite{Fong_2017, fong2018net2vec, petsiuk2018rise, chang2019explaining}. In another line of work, the proposed methods involve training a decoder to estimate the mask, in the same vein as we do in this paper. These works include \cite{dabkowski2017real, Fan2017AdversarialLN, Zolna_2020, phang2020investigating}.   

In the audio domain, notable works on interpretability include \cite{becker2023audiomnist}, which proposed  layer-wise relevance propagation to generate saliency maps over spectrograms. Another noteworthy works include \cite{trinh18_interspeech, kavaki20_interspeech} where authors learn to identify important parts of the input spectrogram by masking additive white noise, within the context of automatic speech recognition. Additionally, \cite{won2019musictagging} proposed creating visualizations using attention layer outputs, while \cite{muckenhirn19_interspeech} suggested using Guided Backpropagation for spectrogram saliency maps.
SLIME \cite{Mishra2017LocalIM, mishra2020reliable} proposes to divide the spectrogram into predefined time/frequency regions (akin to the superpixels in LIME \cite{ribeiro2016why} for images) and determines the feature importance for each region. AudioLIME \cite{haunschmid2020audiolime, chowdhury2021tracing}, on the other hand, defines the LIME superpixels as sources extracted from the input audio and determines a saliency score for each source. More recently \cite{l2i} proposed Listen-to-Interpret (L2I) to learn the classifier relevance for Non-Negative Matrix factorization (NMF) \cite{Lee1999LearningTP} dictionaries, via a decoder trained to estimate NMF activations. This work is particularly relevant to our method, as we also train a decoder. Consequently, we include detailed comparisons with L2I in our experiments.


\section{Methodology}
\label{sec:methodology}
The architecture of the proposed L-MAC method is illustrated in Figure \ref{fig:arch}. Starting with the original audio waveform, we compute the (linear) spectrogram, denoted as $X$.
The audio spectrogram $X$ is then processed by a feature extraction block that computes the features needed by the pretrained classifier. Notably, the pretrained model can internally use various input features, such as FBANKs (mel-spectrogram features), making our approach feature-agnostic.
These features are input to the classifier, which then generates predictions.

To generate an interpretation signal that explains these predictions, the latent representations $h$ of the classifier are fed into the decoder\textcolor{black}{, as illustrated in Figure \ref{fig:arch}. A more detailed diagram of how the classifier representations are used is shown in Figure \ref{fig:unet_diag} in Appendix \ref{app:diag}}. The decoder is trained to generate a binary mask $M$, selecting relevant portions of the original spectrogram $X$. Importantly, the decoder is not specifically tuned to the features used by the pretrained classifier as we apply the binary mask to the (linear) spectrogram of the original audio $X$. This feature-agnostic aspect allows L-MAC to maintain listenability. By inheriting the phase from the original audio, we can indeed invert the masked spectrogram and produce a listenable audio signal as an interpretation outcome.
To train the decoder, we feed the masked input into the classifier, and compute the mask loss. 
The objective is to learn a decoder capable of outputting a mask that accurately selects the region of interest in the input spectrogram.
The main components of L-MAC are detailed in the following sub-sections.

\subsection{The Masking Objective}
The masking loss employed in this work draws inspiration from similar objectives described in \cite{dabkowski2017real, Zolna_2020, phang2020investigating}.
The goal is to maximize the confidence of the classification decision for the masked-in portion of the audio while minimizing it for the masked-out portion.
The overall objective function is the following: 
\begin{align}
    &\min_M \alpha \mathcal L_{in}( f(M \odot X), y) \\
     &  \; \; \; \; - \beta \mathcal L_{out} (f( (1-M) \odot X), y ) + R(M), \notag
\end{align}
where $f(.)$ represents the pretrained classifier being interpreted. The term $\mathcal L_{in}$ represents the categorical cross-entropy loss computed when we input the masked input $X \odot M$ to the pretrained classifier.
In contrast to the aforementioned prior works, the categorical cross-entropy is computed using the network decision as the label, denoted as $y=\arg\max_c f_c(X)$, rather than the actual targets.
Our objective is to minimize this term, as we want the masked signal to capture the elements that influenced the decision made by the classifier.

The term $\mathcal L_{out}$ represents the categorical cross-entropy computed for all parts of the input $x$ not selected by the mask $M$. We aim to maximize it, as we want the mask to exclude information relevant to the pretrained classifier. 
Overall, we engage in a optimization problem where the $\mathcal L_{in}$ term tends to encourage larger masks, while the $\mathcal L_{out}$ term encourages smaller ones. The decoder must find a valuable trade-off between these aspects.
Finally, note that $R(M_\theta(h))$ is a regularization term that includes an $l1$-regularizer to promote sparsity in the estimated mask.   

One important aspect of our work is the use of a neural network, specifically the decoder, to estimate the binary mask $M$. This choice is motivated by our observation that neural networks yield more faithful and understandable masks.
After parameterizing the interpretation mask $M$ with a neural decoder, represented by parameters denoted as $\theta$, the corresponding optimization objective becomes the following,
\begin{align}
    &\min_\theta \lambda_{in} \mathcal L_{in}( \log f(M_\theta(h) \odot X), y) \label{eq:aoloss} \\
    - &\lambda_{out} \mathcal L_{out} (\log f( (1-M_\theta(h)) \odot X), y) + R(M_\theta(h)), \notag
\end{align}
where the decoder $M_\theta$ maps an internal representation $h$ of the classifier to a binary mask.

After the initial mask optimization, this framework allows a fine-tuning stage where the interpretation mask is refined to enhance the quality of the interpretations. This is accomplished by adding a term to the regularizer $R(.)$ as follows:
\[
    R(M_\theta(h)) = \lambda_g \| M_\theta(h) \odot X - X \| + \lambda_s \| M_\theta(h) \|_1,
\]
where $\lambda_g$ and $\lambda_s$ are regularization coefficients, and $X$ represents the spectrogram of the original signal. The first term in the regularization encourages the decoder to produce masked representations close to the original inputs, while the second term promotes sparsity in the mask.
In our best configuration, this guidance is applied exclusively during training in a selective manner. Specifically, we only apply it to data items if the mask after the initial stage is highly similar to the binarized target spectrogram $X$. The similarity is measured by calculating the normalized cosine similarity between these two objects. This selective fine-tuning helps prevent steering the masks away from faithful interpretations. Finally, note that in case data augmentation is used, the target is chosen as the clean signal in the fune-tuning stage. 

\subsection{Producing Listenable Explanations}
State-of-the-art audio and speech classifiers often rely on features computed on top of the linear spectrogram, such as FBANKs (mel-spectrograms). These features intelligently compress the frequency axis, often leading to more compact features that enhance performance.
However, these features are non-invertible due to the compression applied to the frequency axis. 

In our pipeline (Figure \ref{fig:arch}), we tackle this challenge by having the decoder $M_\theta(.)$ output a mask for the linear spectrogram $X$ instead of generating a mask for the specific features used by the pretrained classifier. During training, we convert this masked signal $M_\theta(h) \odot X$ back to the feature domain before computing the training loss in Equation \eqref{eq:aoloss}. The audio domain interpretations are obtained by inverting the linear spectrogram using the phase of the original audio waveform $x$, expressed as:
\begin{align}
    x_\text{interpretation} = \text{ISTFT} \left ( ( M_\theta(h) \odot X ) e^{j X_\text{phase} } \right ).
\end{align}


\section{Experiments}

\cemcamera{In our experiments, we evaluate the faithfulness and understandability of the generated interpretations. To achieve this, we considered two setups: i) Classification under in-domain conditions, and ii) Classification under out-of-domain conditions. We utilized the ESC50 dataset \cite{piczak2015dataset} which contains 50 environmental sound classes for both setups. We also provide additional results on the UrbanSound8k dataset \cite{us8k} in Appendix \ref{app:US8k}.}

\subsection{Metrics}
To measure the faithfulness of classifications we have used the following metrics: 

\textbf{Faithfulness on Spectra (FF):} 
This metric is originally introduced in \cite{l2i}, as a way to measure how important is the generated interpretation for a classifier. The metric is calculated by measuring the drop in class-specific logit value, when the masked out portion of the interpretation mask is input to the classifier. This amounts to calculating, 
\begin{align*}
    \text{FF}_n:= f(X_n)_c - f(X_n\odot(1-M_\theta(h)))_c  
\end{align*}

If this metric is large, this signifies that the masked-in portion of the input spectrogram $X$ is highly influential for the classifier decision for class $c$.  We report the average faithfulness over all examples by reporting the average quantity $\text{FF}:= \sum_n \frac{1}{N} \text{FF}_n$. Larger is better. 

\textbf{Average Increase (AI):} 
Average increase, originally proposed in \cite{Chattopadhay_2018}, measures the increase in confidence for the masked-in portion of the interpretation, and it is calculated as follows:
\begin{align*}
    \text{AI}:= \frac{1}{N} \sum_{n=1}^N \textbf{1}_{[f(X_n\odot M_\theta(h)) > f(X_n)_c]} \cdot 100,  
 \end{align*}
where $\mathbf 1_{[.]}$, is an indicator function which returns one if the argument is true, and zero otherwise. For this metric, larger is better. 

\textbf{Average Drop (AD):}
Average drop, originally proposed in \cite{Chattopadhay_2018}, measures how much confidence is lost when the input image is masked, and calculated as follows:
\begin{align*}
    \text{AD}:=\frac{1}{N} \sum_{n=1}^N \frac{\max(0, f(X_n)_c - f(X_n\odot M_\theta(h))_c) }{f(X_n)_c} \cdot 100.
\end{align*}
For this metric, smaller is better. 

\textbf{Average Gain (AG):}
This metric is first proposed in \cite{zhang2023opticam}, and it measures how much confidence is gained after masking the input image. It is calculated as follows: 
\begin{align*}
    \text{AG}:=\frac{1}{N} \sum_{n=1}^N \frac{\max(0,  f(X_n\odot M_\theta(h))_c - f(X_n)_c )}{1-f(X_n)_c} \cdot 100.
\end{align*}

\textbf{Input Fidelity (Fid-In):}
This metric is introduced in \cite{paissan2023posthoc}, and it measures if the classifier outputs the same class prediction on the masked-in portion of the input image. It is defined as, 
\begin{align*}
    \text{Fid-In} = \frac{1}{N} \sum_{n=1}^N \mathbf 1_{ [\arg \max_c f(X_n)_c = \arg \max_c f_c(X_n \odot M_\theta(h)) ]}.
\end{align*}
Larger values are better. 

\textbf{Sparseness (SPS):}
Sparseness measure is introduced in \cite{chalasani2020concise}, and it measures if only values with large predicted saliency contribute to the prediction of the neural network. Larger values  indicate more sparse/concise saliency maps. We use the  implementation from the Quantus library \cite{hedstrom2023quantus}. 

\textbf{Complexity (COMP):}
Complexity metric is introduced in \cite{bhatt2020evaluating}, and this metric measures the entropy of the distribution of contributions from each feature to the attibution. Smaller values indicate less complex interpretations. We again used the implementation from the Quantus library. 

\subsection{Faithfulness Evaluation}

\begin{table*}[t]
\caption{In-domain quantitative evaluation on the ESC50 dataset. Our results reveal that L-MAC consistently achieves significantly higher faithfulness scores (AI, AD, AG, FF, Fid-In) compared to other methods.}
\label{tab:ID-ESC50}
\vskip 0.15in
\centering
\begin{tabular}{ll|ccccccc}
\toprule
&\textbf{Metric} & AI ($\uparrow$) & AD ($\downarrow$) & AG ($\uparrow$) & FF ($\uparrow$) & Fid-In ($\uparrow$) & SPS ($\uparrow$) & COMP ($\downarrow$) \\
\midrule
& {Saliency} & 0.00 & 15.79 & 0.00 & 0.05	& 0.07 & 0.39	& 5.48   \\ 
 &{Smoothgrad} & 0.00	&15.71 & 0.00 & 0.03 &0.05	&0.42	&5.32 \\
 &IG & 0.25 & 15.45 & 0.01 &0.07 & 0.13& 0.43 & 5.11 \\
&GradCAM & 8.50 & 10.11 & 1.47 &	0.17 &	0.33 &	0.34 &	5.64 \\
&Guided GradCAM & 0.00 & 15.61 &0.00& 0.05 &	0.06 &	0.44 & 5.12 \\
 \vbox{\hbox{\multirow{2}{*}{\rotatebox[origin=c]{90}{STFT-Mel}}}\vspace{6pt}} &Guided Backprop & 0.00	& 15.66 &0.00&0.05 & 0.06	 & 0.39 &	5.47 \\
& L2I, RT=0.2	& 1.63	&12.78 &	0.42	&0.11	&0.15&	0.25	&5.50 \\
&L2I, RT=0.4	&1.13&	11.72&	0.15&	0.08&	0.11&	0.23&	4.41 \\
&L2I, RT=0.60&	0.50&	7.90&	0.05&	0.04 & 0.06&	0.14 &	2.61 \\
&L2I, RT=0.80&	0.13	&3.21&	0.01&	0.01&	0.02&	0.06&	\textbf{0.98} \\
&{SHAP} & 0.00	&15.79 &	0.00 &	0.05 &	0.06 &	0.43 & 5.24 \\
&\textbf{L-MAC (ours)}  & \textbf{36.25} &	\textbf{1.15}	& \textbf{23.50}	&0.20 &	\textbf{0.42}	& \textbf{0.47} &	4.71 \\ 
&{L-MAC, FT, $\lambda_g=4$ (ours)}  & 32.37 &	1.98	&18.74	& \textbf{0.21} &	0.41 &	0.43	& 5.20 \\
&L-MAC, FT, $\lambda_g=16$ (ours) & 27.12 & 3.32 & 16.18& 0.19 &	0.39 &	0.44 &	5.03  \\
&L-MAC, FT, $\lambda_g=32$ (ours)	& 23.00 & 4.42 & 12.63 & 0.18 & 0.37 &	0.45 &	4.92 \\

\midrule
&Saliency &0.00	&15.81	&0.00	&0.10&	0.07&	0.39&	4.53 \\
&Smoothgrad &0.00	&15.61	&0.00	&0.07&	0.04&	0.39&	4.54 \\
&IG &0.00	&15.55&	0.00&	0.12&	0.08&	0.42&	4.36 \\
 \vbox{\hbox{\multirow{2}{*}{\rotatebox[origin=c]{90}{Mel}}}\vspace{-2pt}} &GradCAM &	7.00&	10.93&	1.04&	0.17&	0.29&	0.34&	\textbf{4.72}\\
&Guided GradCAM&	0.125&	15.40&	6.67&	0.08&	0.07&	\textbf{0.45}&	4.17\\
&Guided Backprop&	0.125&	15.54&	0.00&	0.10&	0.08&	0.39&	4.53 \\
&SHAP&	0.00&	15.57&	0.00&	0.11&	0.08&	0.41&	4.42 \\
&\textbf{L-MAC (ours)} &	35.63&	1.59&	\textbf{24.28}&	0.22&	\textbf{0.42}& \textbf{0.45}&	4.11 \\
&\textbf{L-MAC (ours)} FT, $\lambda_g = 4$ & \textbf{36.13} & \textbf{1.28}& 21.15& \textbf{0.23}& \textbf{0.42}& 0.32& 4.71\\
\bottomrule
\end{tabular}
\vspace{-0.3cm}
\end{table*}

\begin{table*}[t]
\caption{Out-of-Domain Quantitative Evaluation for the ESC50 Dataset. In out-of-distribution conditions, L-MAC consistently outperforms other methods across all evaluated metrics.}
\label{tab:OOD-ESC50}
\vskip 0.15in

\centering
\resizebox{.9\textwidth}{!}{
\begin{tabular}{ll|ccccccc}
\toprule
&\textbf{Metric} & AI ($\uparrow$) & AD ($\downarrow$) & AG ($\uparrow$) & FF ($\uparrow$) & Fid-In ($\uparrow$) & SPS ($\uparrow$) & COMP ($\downarrow$) \\
\midrule
& Saliency 	& 0.62 &	31.73	&0.07 & 0.06 &	0.12 & 0.76 & 11.06 \\
& Smoothgrad &0.12 &31.84 & 0.00 &0.06 &0.13 & 0.83 & 10.66 \\ 
 & IG  & 0.37  & 31.15 &	0.03 &	0.12 & 0.26 &	0.87 & 10.22  \\
& L2I	&5.00	&25.65	&1.00 &	0.20 &	0.35 & 0.52 &	10.99 \\
 \vbox{\hbox{\multirow{2}{*}{\rotatebox[origin=c]{90}{STFT-Mel}}}\vspace{-6pt}}& GradCAM &	14.12	& 17.62 & 7.46 & 0.25	& 0.00 & 0.91 & 9.66 \\
&Guided GradCAM & 0.00	& 31.74 & 0.00 & 0.07	& 0.11	& 0.89 & 10.24 \\
&Guided Backprop & 0.63 & 31.73 & 0.07 & 0.06 & 0.11 & 0.76 & 11.06\\
&SHAP & 0.00 & 31.81 &	0.00 & 0.07 &	0.14 & 0.84 & 10.58 \\
& \textbf{L-MAC (ours)} & \textbf{60.63} & \textbf{4.82} & \textbf{35.85} & \textbf{0.39} &	\textbf{0.81} &	\textbf{0.94} & \textbf{9.61} \\
& L-MAC FT, $\lambda_g=4$ (ours) & 50.75 &6.73 & 26.00 & \textbf{0.39} & 0.78 &	0.84 &10.51 \\
& L-MAC  FT, $\lambda_g=16$ (ours) & 37.62 & 10.67 & 19.29 & 0.34 &	0.70 & 0.87 & 10.19 \\
& L-MAC - FT, $\lambda_g = 32$ (ours) & 28.88 & 12.69 & 14.56 & 0.32 & 0.66 &	0.89 & 10.01 \\
& L-MAC - FT, $\lambda_g = 4$ (ours), CCT 0.7 & 52.87	 & 6.71	& 29.46& 0.38 &0.78	&0.93	& 9.76 \\
& L-MAC - FT, $\lambda_g=16$ (ours), CCT 0.7	& 45.87 & 8.12 & 23.91 & 0.37 & 0.74 & 0.91 & 9.93 \\
& L-MAC - FT, $\lambda_g=32$ (ours), CCT 0.7 & 38.50	& 9.62 & 19.11 & 0.35 &	0.70 &	0.89 & 10.04 \\

\midrule
& Saliency	& 0.38 & 31.64 & 0.01 & 0.15 &	0.12 & 0.77 & 9.17  \\
&Smoothgrad & 0.25 & 31.66 &	0.01 &0.14 & 0.11 &	0.79 & 9.03 \\
&IG & 0.12 & 31.52 & 0.01 &	0.19 &	0.19 &	0.84 &	8.62 \\
&GradCAM & 19.88 & 18.85 &	4.67 & 0.34 & 0.69 & 0.66 &	9.49 \\
 \vbox{\hbox{\multirow{2}{*}{\rotatebox[origin=c]{90}{Mel}}}\vspace{-6pt}}&Guided GradCAM	& 0.00	 & 31.68 &	0 & 0.14 & 0.12 & 0.89 & 10.24 \\
&Guided Backprop & 0.38 & 31.64 & 0.01 & 0.15 &	0.12 & 0.77 & 9.16 \\
&SHAP & 0.25 & 31.60 & 0.00 & 0.17 & 0.15 & 0.82 &	8.81 \\
&\textbf{L-MAC (ours)} & 60.25 & \textbf{4.84} &	\textbf{34.72} &	\textbf{0.44} & 0.80	& \textbf{0.90} & \textbf{8.29} \\
& \textbf{L-MAC - FT, $\lambda_g=4$ (ours)} & \textbf{60.75} & \textbf{4.84} &	29.34 &	\textbf{0.44} & \textbf{0.83} & 0.64 & 9.38 \\
& L-MAC - FT, $\lambda_g=16$ (ours) & 45.75 & 9.93 & 17.04 & 0.43 &	0.80 & 0.69 & 9.16 \\
& L-MAC - FT, $\lambda_g=32$	(ours) & 37.50	& 8.65 & 14.08 & 0.43 &	0.77 & 0.70 & 9.05 \\
\bottomrule
\end{tabular}
}
\vspace{-0.4cm}

\end{table*}

\label{sec:esc50quant}

In these experiments, we first train a CNN14 classifier \cite{kong2020panns} on the ESC-50 dataset \cite{piczak2015dataset} augmented with WHAM! noise, to simulate real-world mixtures. The classifier is trained on folds 1, 2, and 3 and obtains 75\% and 78\% classification accuracy on folds 5 and 4, respectively. The CNN14 classifier we employed has 12 2D convolutional layers and is pre-trained on the VGG-sound dataset \cite{Chen2020vggsound} using SimCLR \cite{chen2020simple}. The decoder of L-MAC consists of a series of transposed 2D convolutional layers. Each convolutional layer upsamples the time and frequency axis. The classifier's representations are fed to the decoder in a U-Net-like fashion to incorporate information at different time-frequency resolutions (as shown in Figure \ref{fig:unet_diag} in Appendix \ref{app:diag}). Specifically, the decoder takes the four deepest representations of the classifier.

We then freeze the weights of the classifier and train a decoder on the same training set as the classifier (ESC50 + WHAM! noise), as shown in Figure \ref{fig:arch}. We perform two sets of evaluations on the ESC-50 dataset to validate the robustness of L-MAC in real-world settings. First, we evaluate L-MAC on in-domain data, where the interpreter is tested against data similar to the one in the training set (ESC 50 folds 4 and 5 with WHAM! noise). The results are provided in Table \ref{tab:ID-ESC50}. In the second place, we evaluate L-MAC performance on out-of-domain data. We generate out-of-domain data by creating mixtures of samples from folds 4 and 5. The results are provided in Table \ref{tab:OOD-ESC50}. We report results obtained with STFT and Mel domain spectra for in-domain and out-of-domain data. \cemcamera{We have also reported additional results on the UrbanSound8k dataset \cite{us8k} in Table \ref{tab:OOD-ESC50}, in Appendix \ref{app:US8k}.}

To compare L-MAC with the literature, we used several gradient-based methods such as standard saliency maps \cite{simonyan2014deep}, SmoothGrad \cite{smilkov2017smoothgrad}, IntegratedGradients \cite{sundararajan2017axiomatic}, GuidedBackProp \cite{springenberg2015striving}, and decoder based audio specific method, Listen-to-Interpret (L2I) \cite{l2i}, and we also include SHAP \cite{lundberg2017unified}. For L2I, we have reported results using relevance thresholds RT=0.2, 0.4, 0.6, and 0.8 (an important hyperparameter for the L2I method). We have used the Captum implementations \cite{kokhlikyan2020captum} for the gradient-based methods and SHAP and adapted the SpeechBrain \cite{ravanelli2021speechbrain} implementation for L2I. For L-MAC, we have obtained results for finetuning strengths $\lambda_g=4,16,32$. \textcolor{black}{The implementation of our experimental setup is accessible through the companion website.\footnote{\url{https://fpaissan.github.io/lmac}}}.

For in-domain data, we observe in Table \ref{tab:ID-ESC50} that L-MAC generally results in better faithfulness scores (AI, AD, AG, FF, Fid-In) compared to the baselines. Since we work with a classifier trained in the Mel domain, we would like to note that we evaluated the methods in cases where we both mask in the STFT domain (denoted with STFT-Mel in Tables \ref{tab:ID-ESC50}, \ref{tab:OOD-ESC50}) and in the Mel domain (denoted with Mel in Tables \ref{tab:ID-ESC50}, \ref{tab:OOD-ESC50}). This is important, as masking in the STFT domain gives the ability to produce listenable interpretations. We observe that for L-MAC, the STFT domain interpretations result in similar faithfulness scores compared to the Mel domain. More specifically, for the in-domain evaluation, we observe that the gradient-based methods such as standard saliency, Smoothgrad, IG, GradCAM, Guided BackProp, and SHAP generally result in less faithful interpretations than the decoder-based methods such as L2I and L-MAC. We see that the most faithful results are obtained without having the finetuning (FT) stage in L-MAC. However, with additional finetuning, we see that L-MAC can produce results that only lose slightly from their faithfulness scores while increasing the understandability of the interpretation, as shown by user preference. We observe that after finetuning, L-MAC can still be more faithful compared to the other baselines we investigate, including our L2I implementation. We also see that in terms of the sparsity metric, L-MAC has better overall numbers than L2I and the other gradient-based methods. In terms of the complexity metric, it is comparable with the gradient-based methods and with L2I with a relevance threshold of 0.2. Note that with larger relevance threshold values, L2I returns less active interpretations. 



In Table \ref{tab:OOD-ESC50}, we show the metrics on out-of-distribution data. In this case, the gradient-based posthoc saliency methods also do not yield very faithful results except for GradCAM. We also observe that L-MAC outperforms L2I even if a severe finetuning strength of $\lambda_g = 32$ is employed. Another observation is that the listenable STFT version of L-MAC yields comparable faithfulness results to the direct Mel domain interpretations yielded by L-MAC. For this data, we also try increasing the cross-correlation threshold (denoted with CCT in Table \ref{tab:OOD-ESC50}) between the interpretations and the target mask during training (as discussed in Section \ref{sec:methodology}). We see that using a larger CCT generally helps increase the interpretations' faithfulness for larger $\lambda_g$ values.

\subsection{Qualitative Evaluation}

\begin{figure*}
    \centering
    \includegraphics[width=0.90\textwidth, trim=0cm 0cm 0cm 0cm, clip]{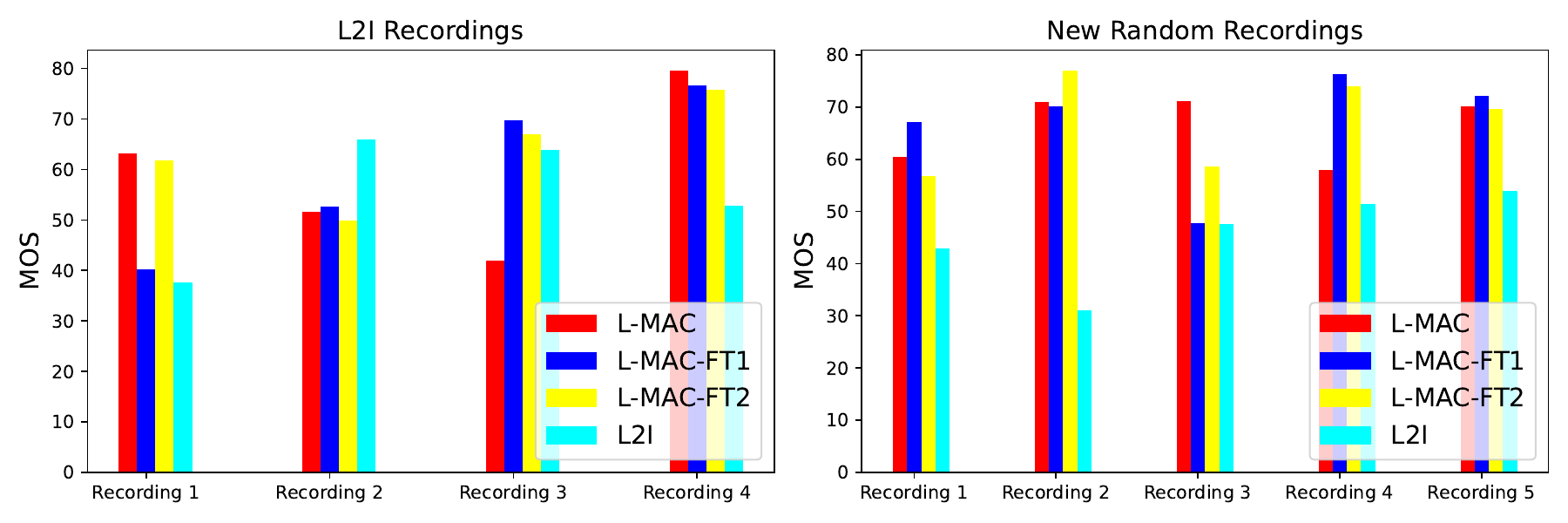}
    \caption{The Mean Opinion Scores (MOS) obtained in the user study. \textbf{(Left)} MOS values obtained on recordings from L2I companion website \textbf{(Right)} MOS values obtained on newly created random recordings with two sound classes. }
    \label{fig:esc50qualitative}
    \vspace{-0.5cm}
\end{figure*}

\tikzstyle{dictsmall} = [draw, thick, fill=white!10, rectangle, 
    minimum height=1.0cm, minimum width=5cm] 
\begin{figure*}[ht]
    \centering
    \resizebox{0.99\textwidth}{!}{
    \centering
    \begin{tikzpicture}[auto] 
        \node [draw=none, fill=none] (ex)  { \includegraphics[width=0.4\textwidth]{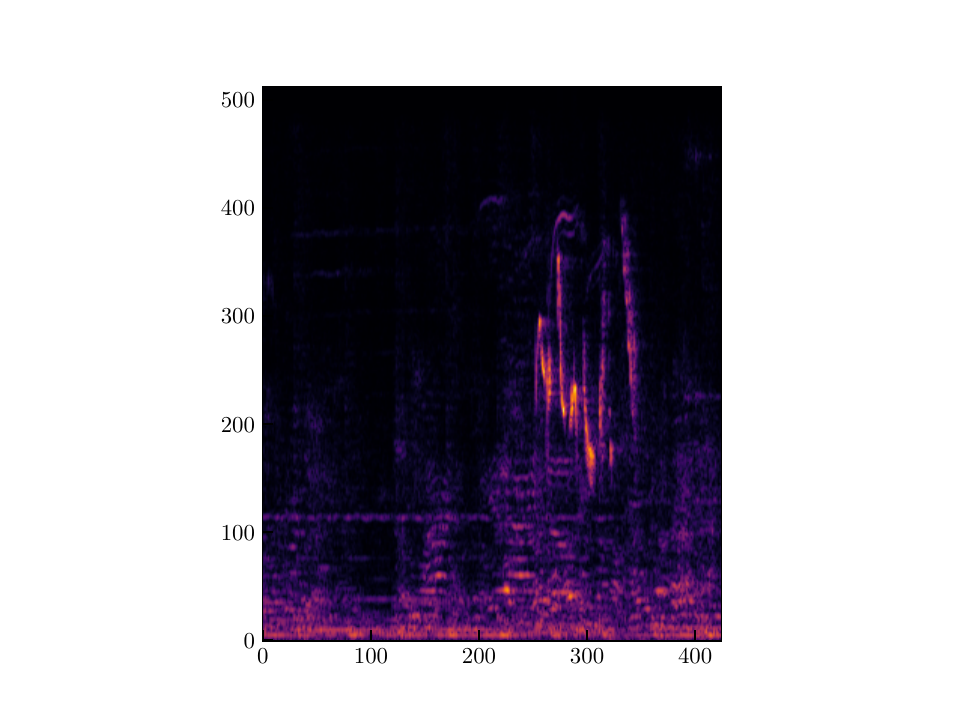} };
        \node [draw=none, fill=none, left of=ex, xshift=-1.2cm] (label)  { \rotatebox{90}{\Large L-MAC}};
        \node [draw=none, fill=none, right of=ex, xshift=4cm] (ex1)  { \includegraphics[width=0.4\textwidth]{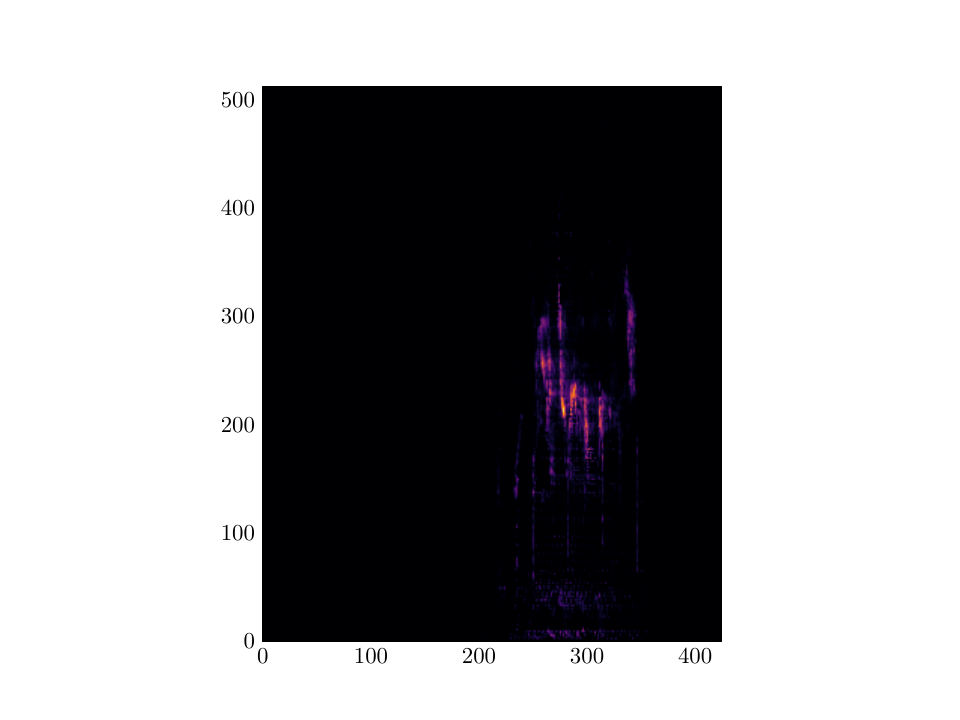} };
        \node [draw=none, fill=none, right of=ex1, xshift=4cm] (ex2)  { \includegraphics[width=0.4\textwidth]{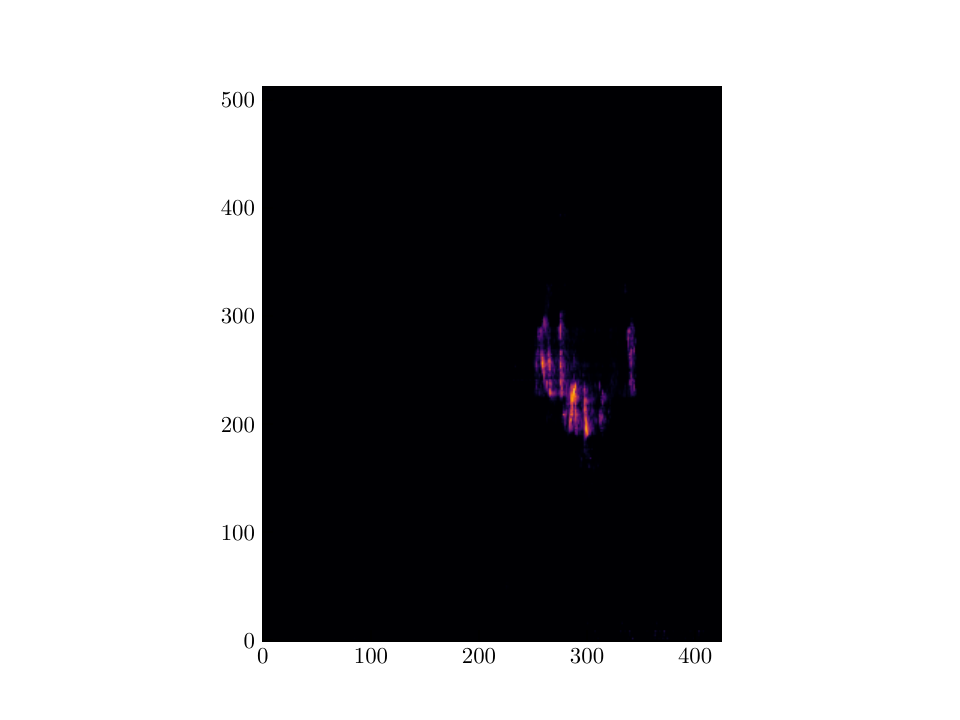} };
        \node [draw=none, fill=none, right of=ex2, xshift=4cm] (ex3)  { \includegraphics[width=0.4\textwidth]{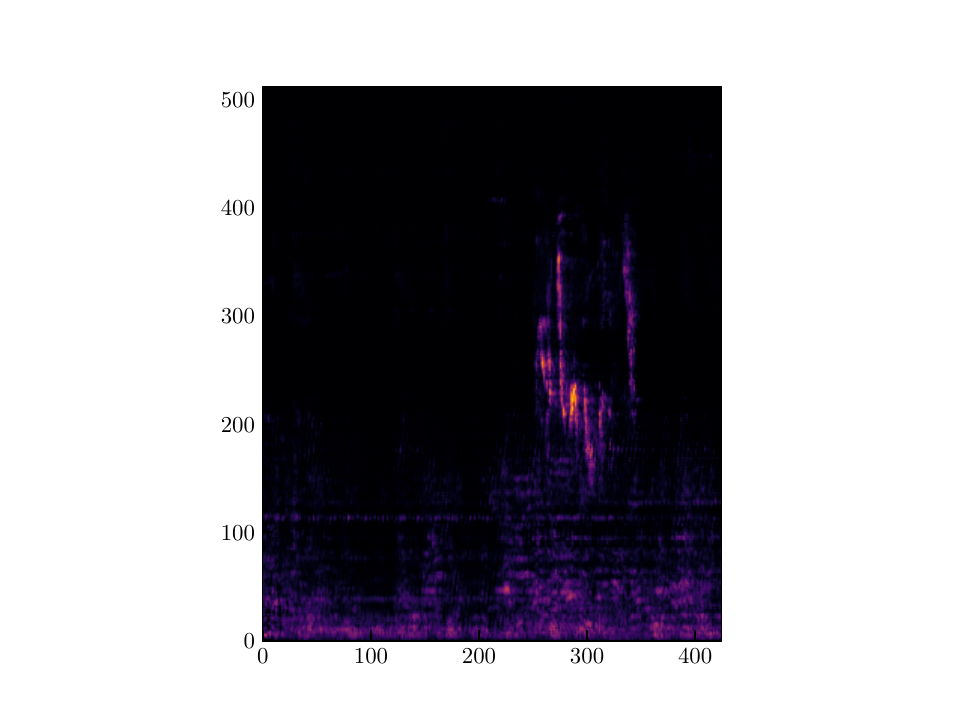} };
        \node [draw=none, fill=none, right of=ex3, xshift=4cm] (ex4)  { \includegraphics[width=0.4\textwidth]{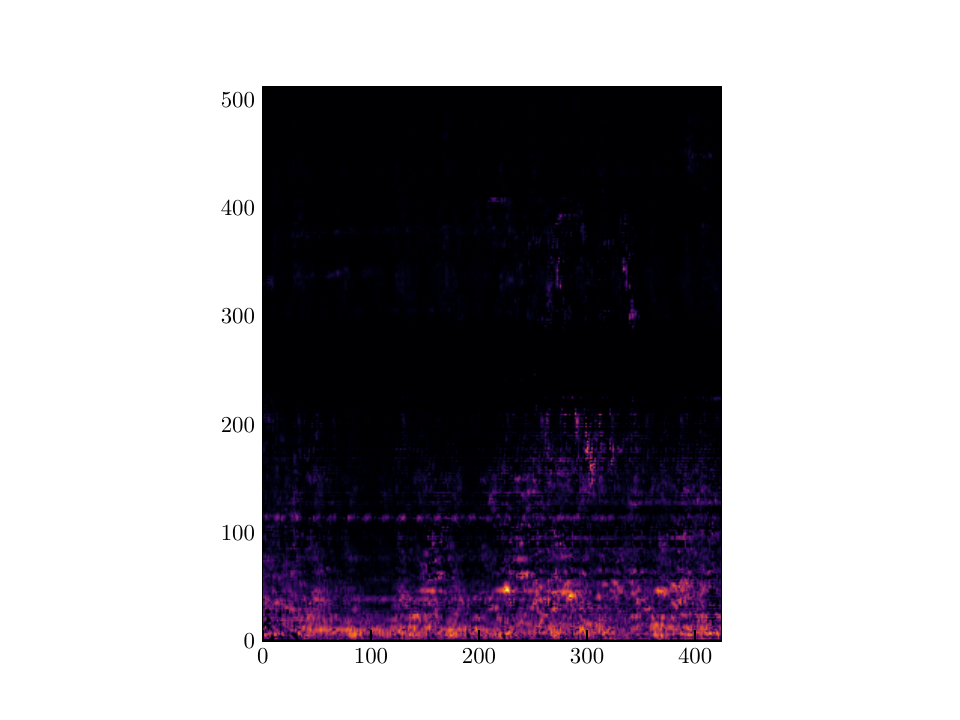} };
        \node [draw=none, fill=none, right of=ex4, xshift=4cm] (ex5)  { \includegraphics[width=0.4\textwidth]{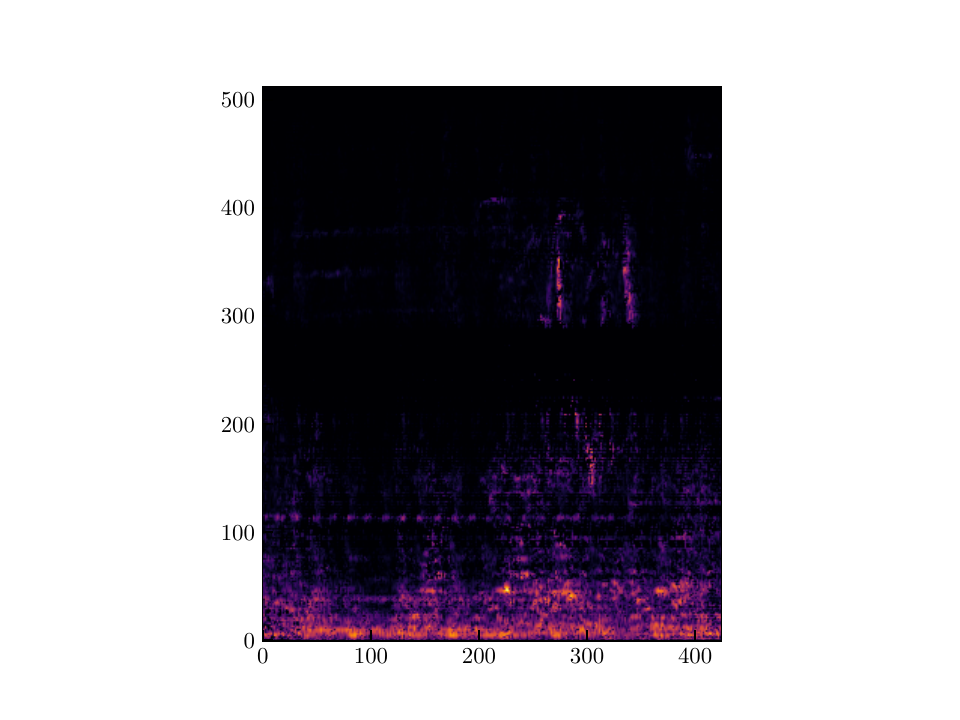} };
        \node [draw=none, fill=none, right of=ex5, xshift=4cm] (ex6)  { \includegraphics[width=0.4\textwidth]{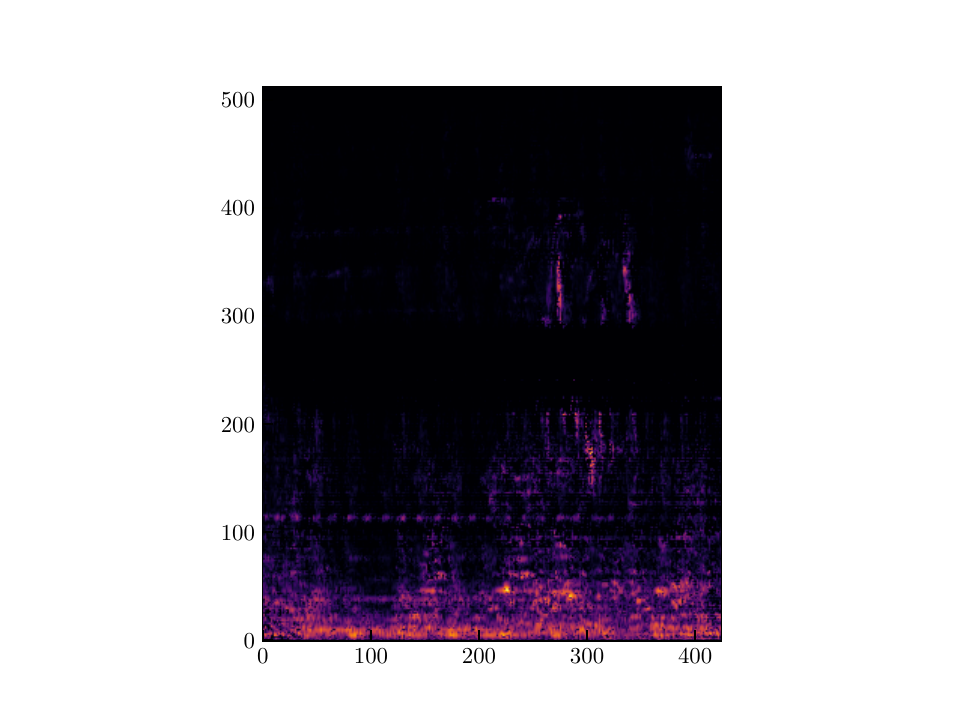} };
        \node [draw=none, fill=none, right of=ex6, xshift=4cm] (ex7)  { \includegraphics[width=0.4\textwidth]{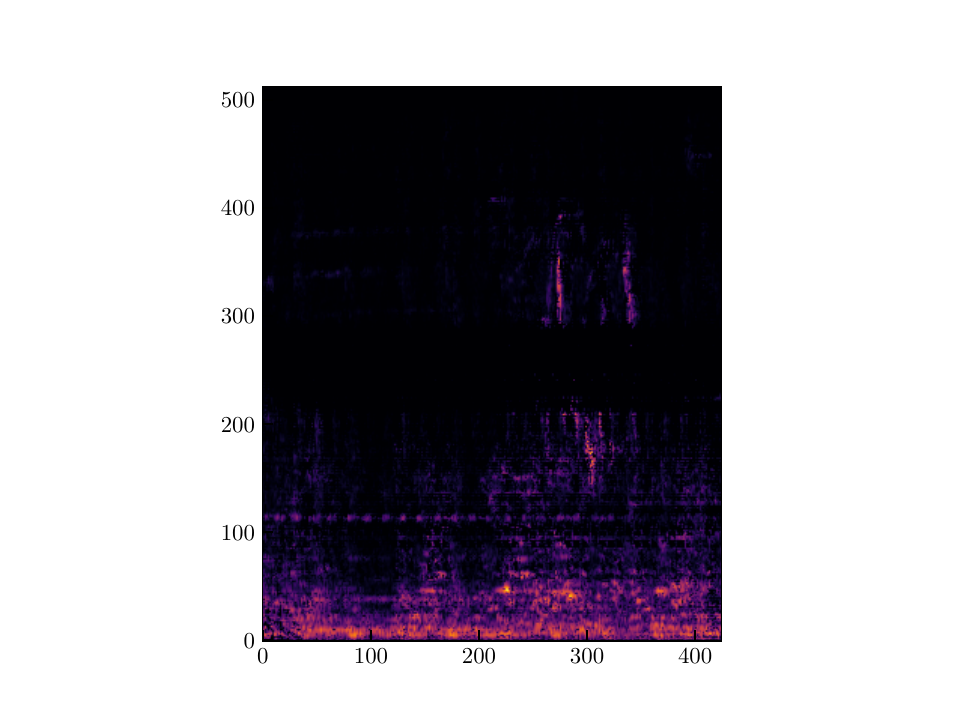} };

        \node [draw=none, fill=none, above of=ex2, yshift=1.2cm] (start)  { };
        \node [draw=none, fill=none, above of=ex6, xshift=4.5cm, yshift=1.2cm] (end)  { };
        \draw[->, thick, scale=3, line width=1.5pt] (start.north west) -- (end.north east) node[midway, above] {\Huge {Cascading randomization (Conv Block ID)}};
        
    \end{tikzpicture}
    }
    \vspace{-0.2cm}
    \resizebox{0.99\textwidth}{!}{
    \centering
    \begin{tikzpicture}[auto] 
        \node [draw=none, fill=none] (ex)  { \includegraphics[width=0.4\textwidth]{icml2024/mrt_viz/ft_4/original.pdf} };
        \node [draw=none, fill=none, left of=ex, xshift=-1.2cm] (label)  { \rotatebox{90}{\Large L-MAC FT4}};
        \node [draw=none, fill=none, right of=ex, xshift=4cm] (ex1)  { \includegraphics[width=0.4\textwidth]{icml2024/mrt_viz/ft_4/qualitative_ft_4.pdf} };
        \node [draw=none, fill=none, right of=ex1, xshift=4cm] (ex2)  { \includegraphics[width=0.4\textwidth]{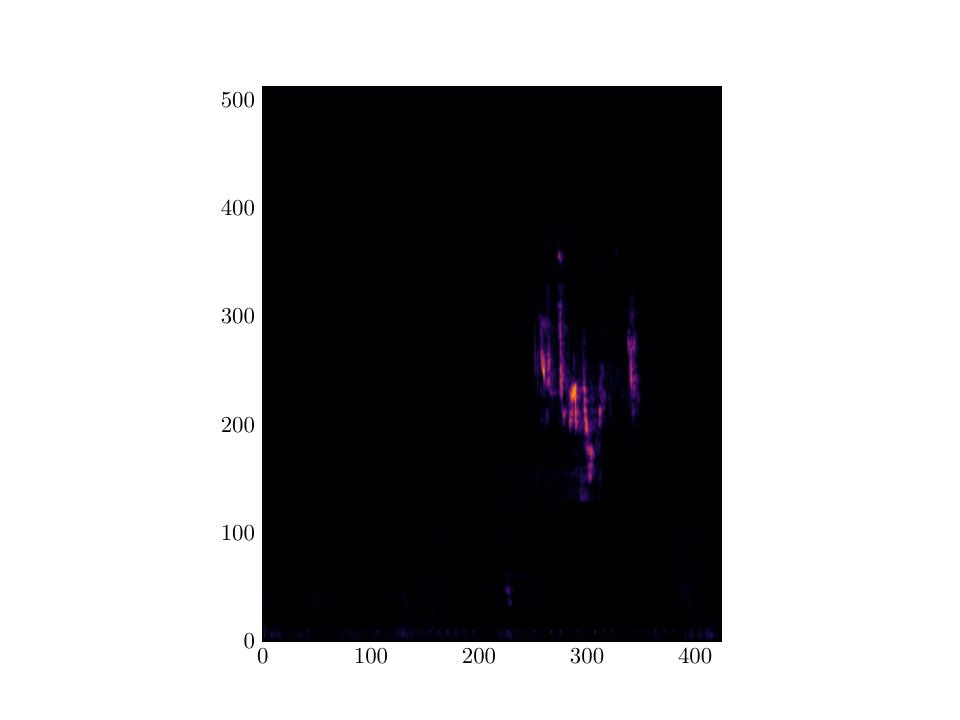} };
        \node [draw=none, fill=none, right of=ex2, xshift=4cm] (ex3)  { \includegraphics[width=0.4\textwidth]{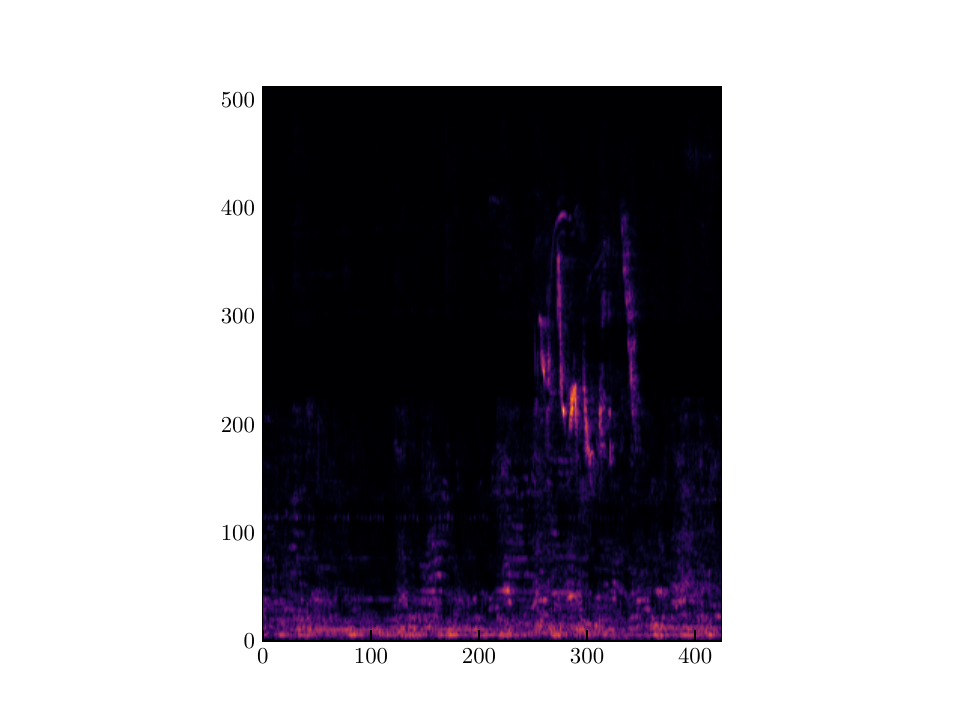} };
        \node [draw=none, fill=none, right of=ex3, xshift=4cm] (ex4)  { \includegraphics[width=0.4\textwidth]{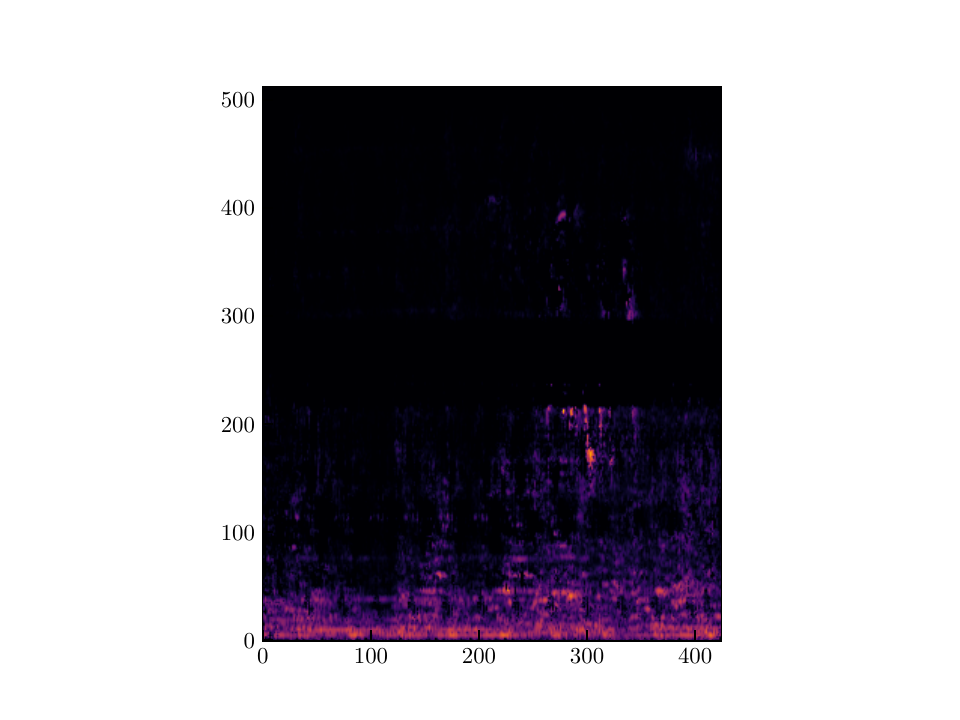} };
        \node [draw=none, fill=none, right of=ex4, xshift=4cm] (ex5)  { \includegraphics[width=0.4\textwidth]{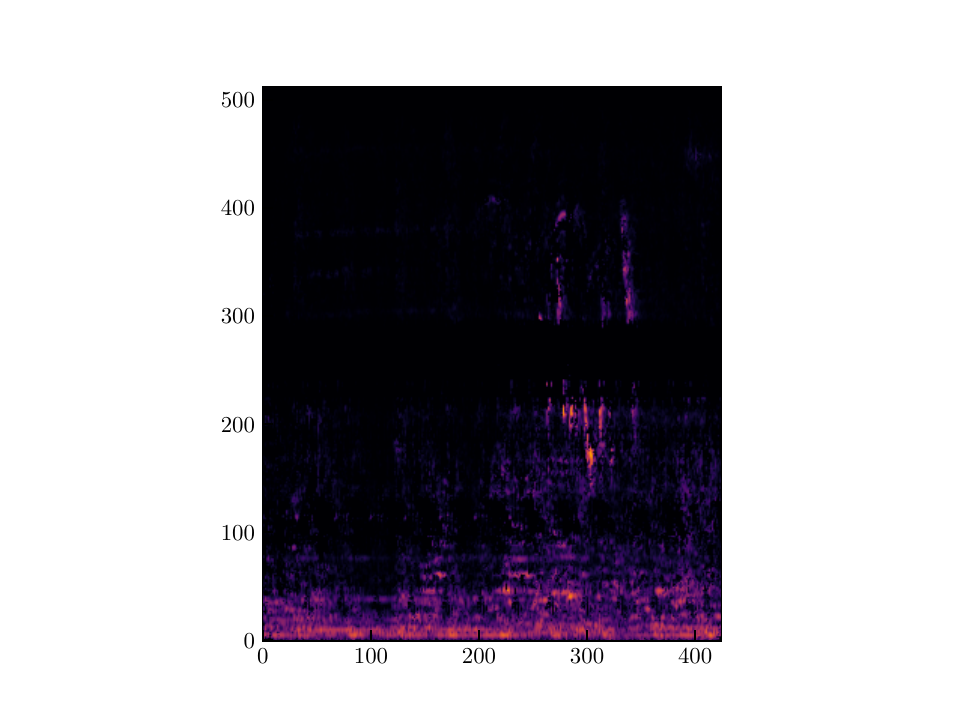} };
        \node [draw=none, fill=none, right of=ex5, xshift=4cm] (ex6)  { \includegraphics[width=0.4\textwidth]{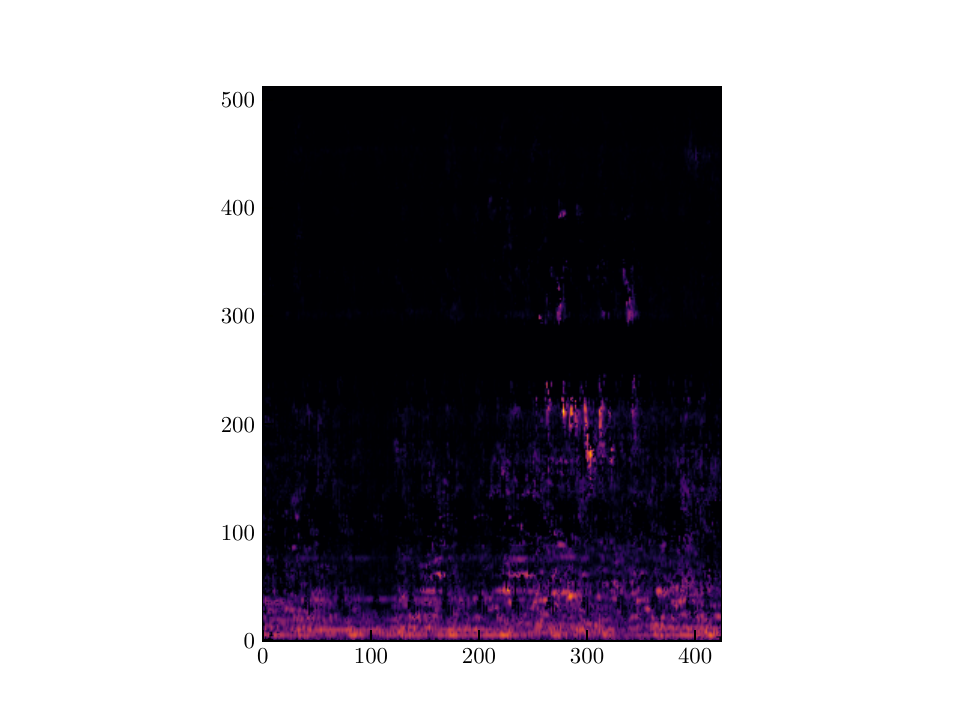} };
        \node [draw=none, fill=none, right of=ex6, xshift=4cm] (ex7)  { \includegraphics[width=0.4\textwidth]{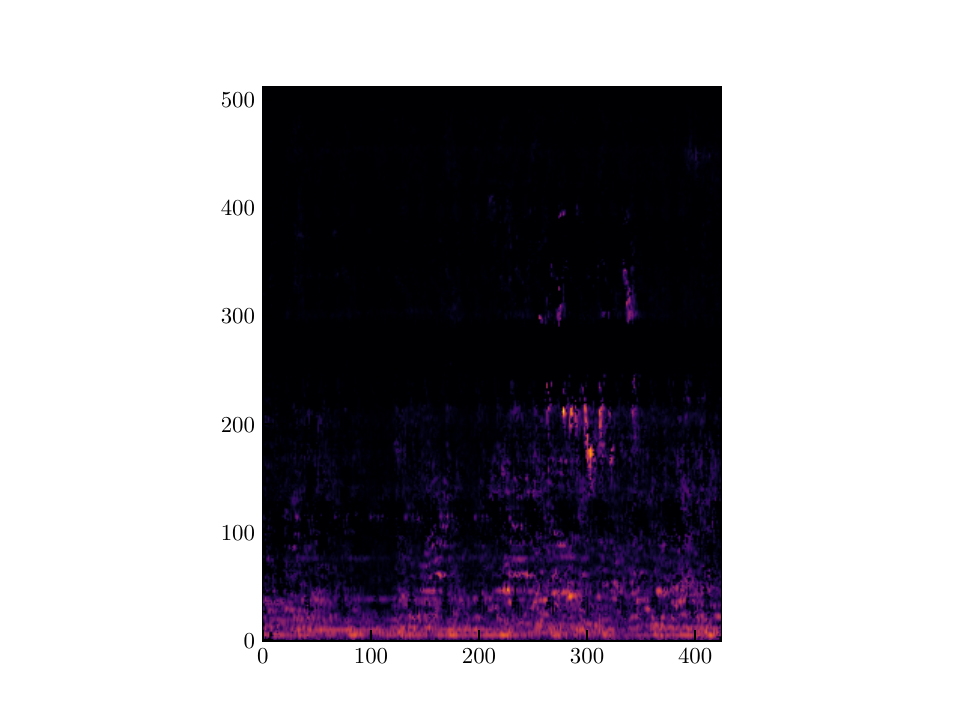} };

    \end{tikzpicture}
    }
        \resizebox{0.99\textwidth}{!}{
    \centering
    \begin{tikzpicture}[auto] 
        \node [draw=none, fill=none] (ex)  { \includegraphics[width=0.4\textwidth]{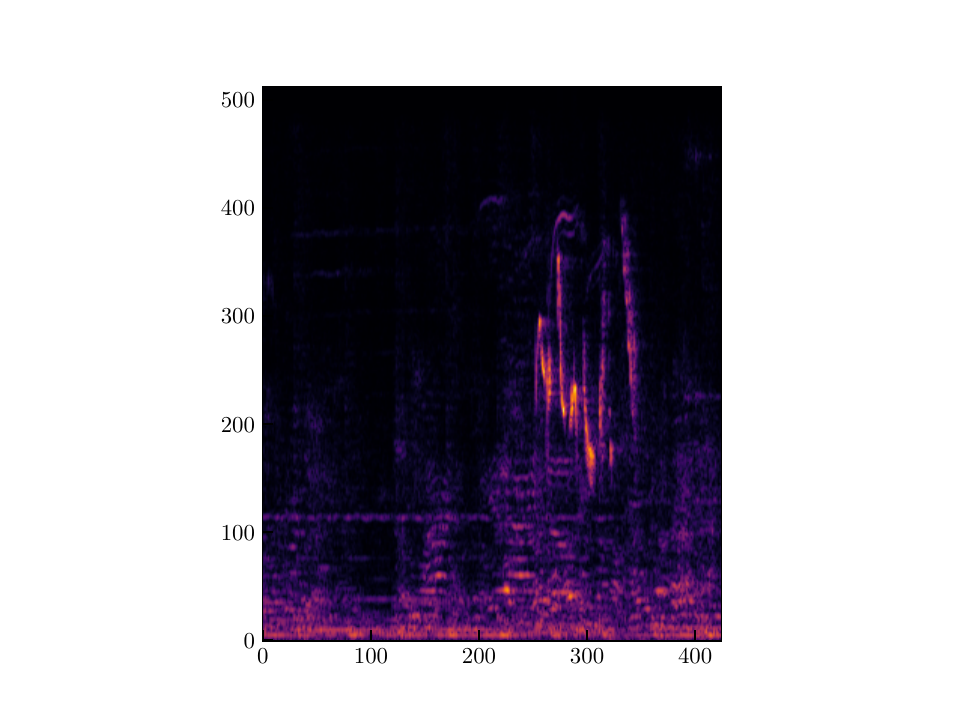} };
        \node [draw=none, fill=none, left of=ex, xshift=-1.2cm] (label)  { \rotatebox{90}{\Large L-MAC FT4-CCT07}};
        \node [draw=none, fill=none, right of=ex, xshift=4cm] (ex1)  { \includegraphics[width=0.4\textwidth]{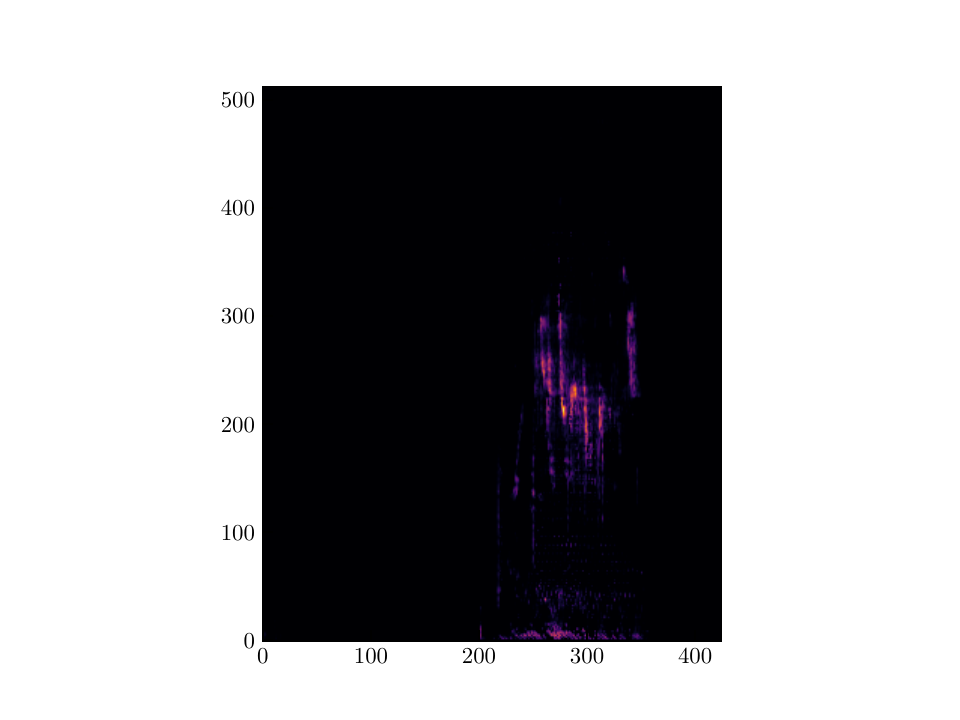} };
        \node [draw=none, fill=none, right of=ex1, xshift=4cm] (ex2)  { \includegraphics[width=0.4\textwidth]{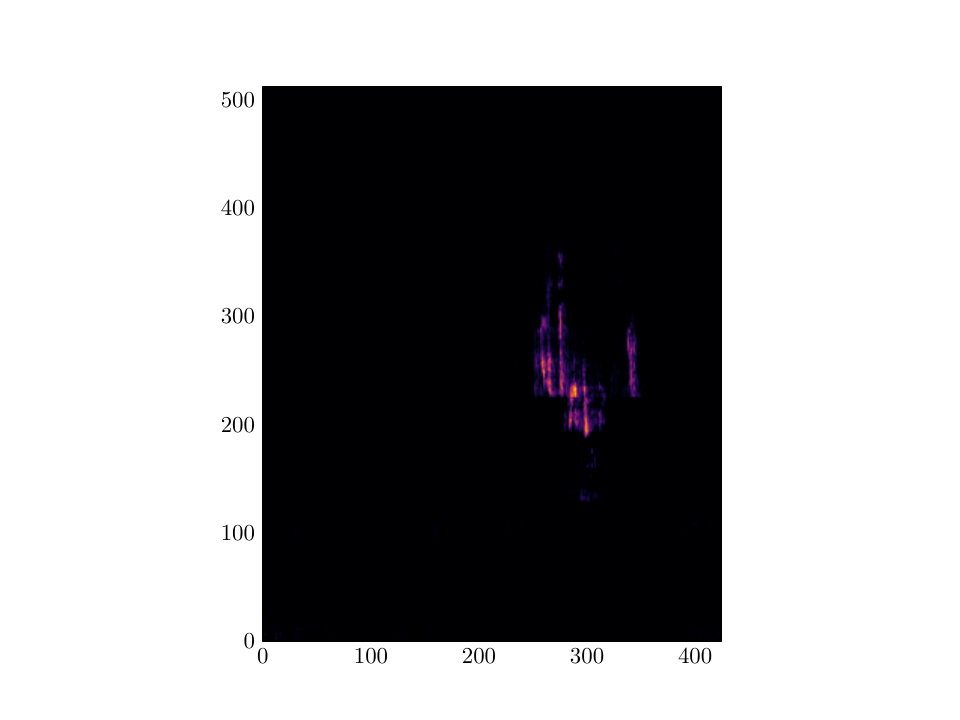} };
        \node [draw=none, fill=none, right of=ex2, xshift=4cm] (ex3)  { \includegraphics[width=0.4\textwidth]{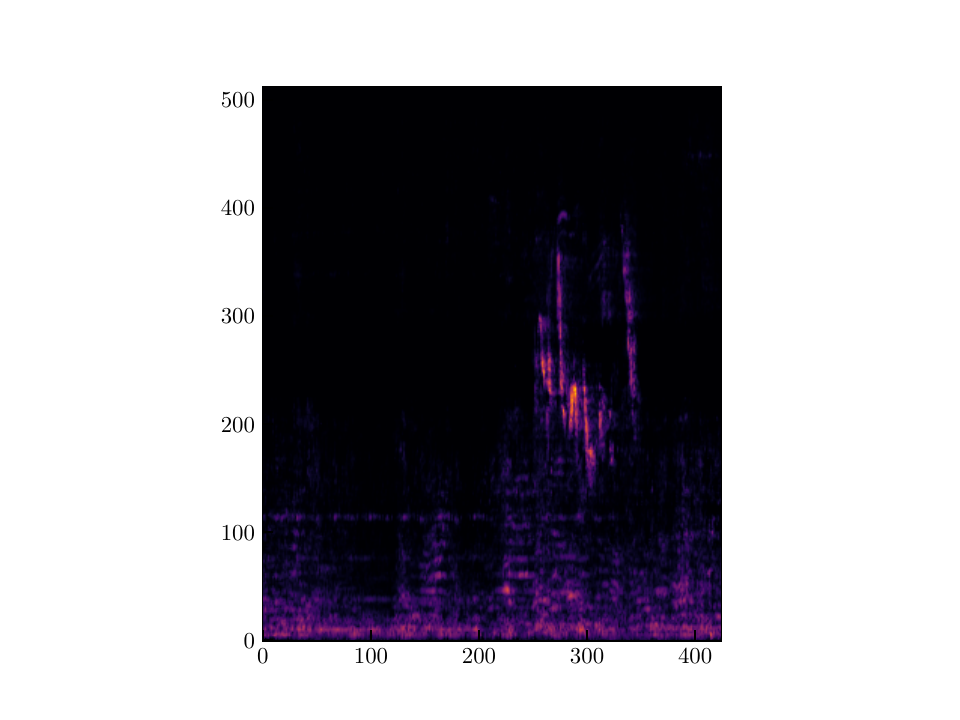} };
        \node [draw=none, fill=none, right of=ex3, xshift=4cm] (ex4)  { \includegraphics[width=0.4\textwidth]{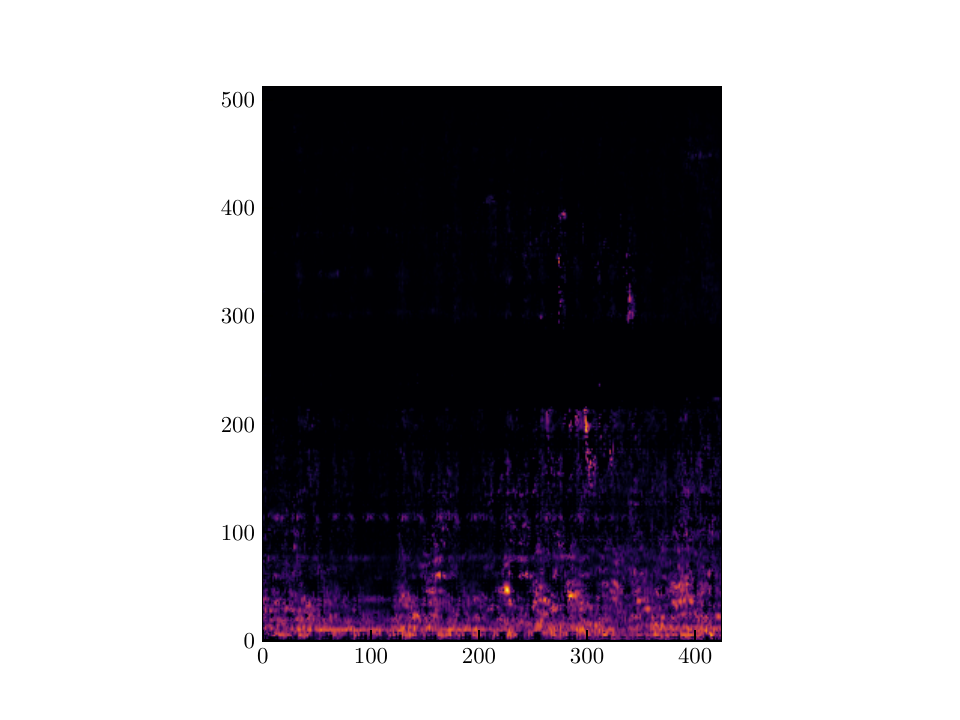} };
        \node [draw=none, fill=none, right of=ex4, xshift=4cm] (ex5)  { \includegraphics[width=0.4\textwidth]{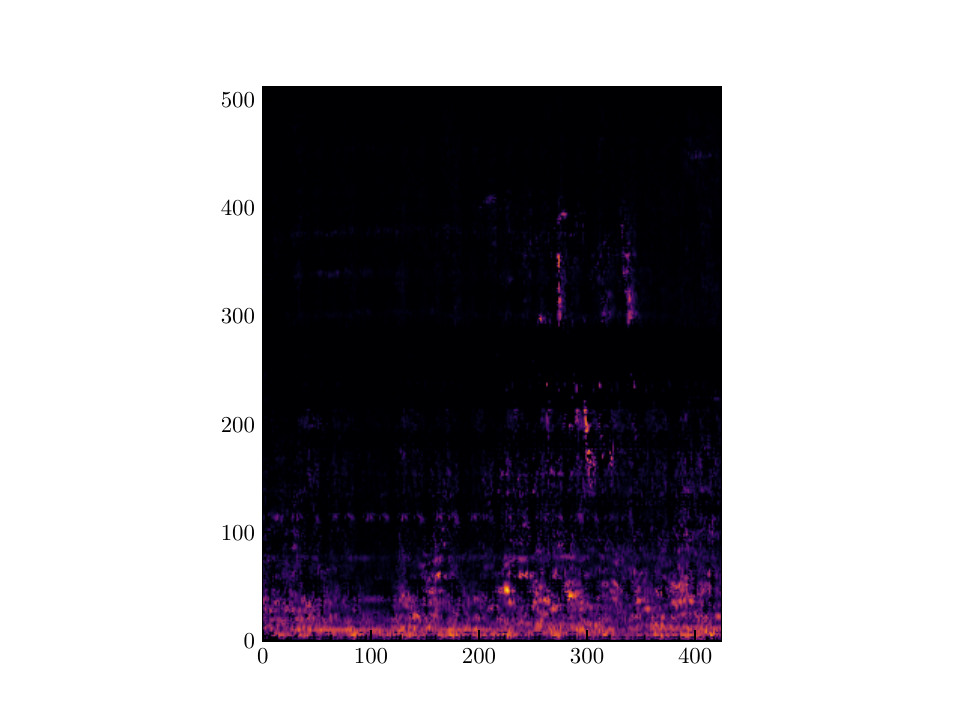} };
        \node [draw=none, fill=none, right of=ex5, xshift=4cm] (ex6)  { \includegraphics[width=0.4\textwidth]{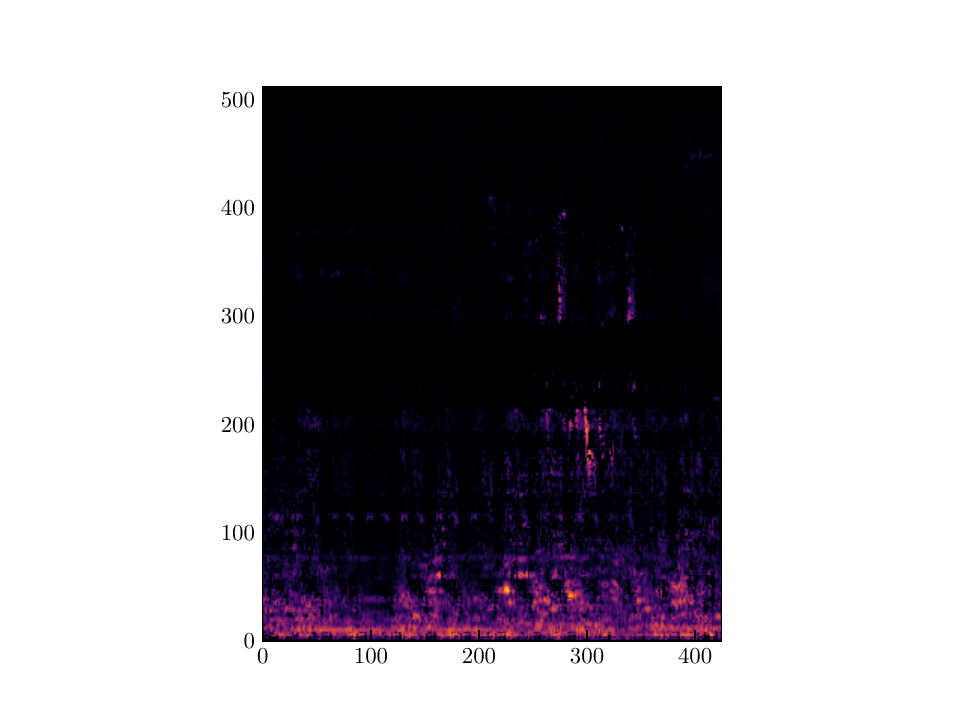} };
        \node [draw=none, fill=none, right of=ex6, xshift=4cm] (ex7)  { \includegraphics[width=0.4\textwidth]{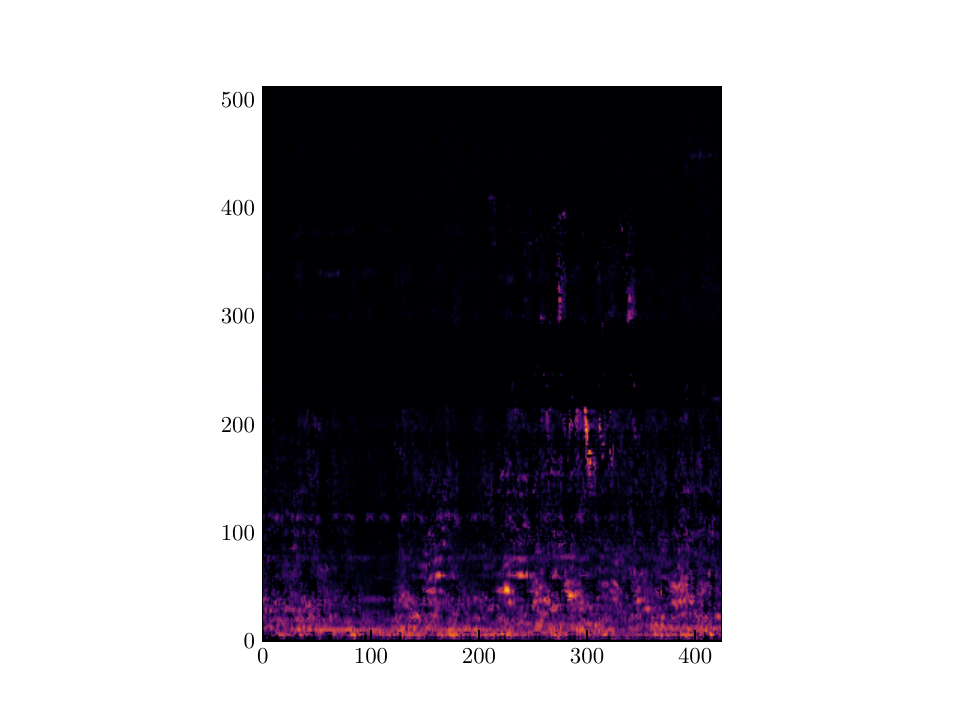} };
    \end{tikzpicture}
    }
       \resizebox{0.99\textwidth}{!}{
    \centering
    \begin{tikzpicture}[auto] 
        \node [draw=none, fill=none] (ex)  { \includegraphics[width=0.4\textwidth]{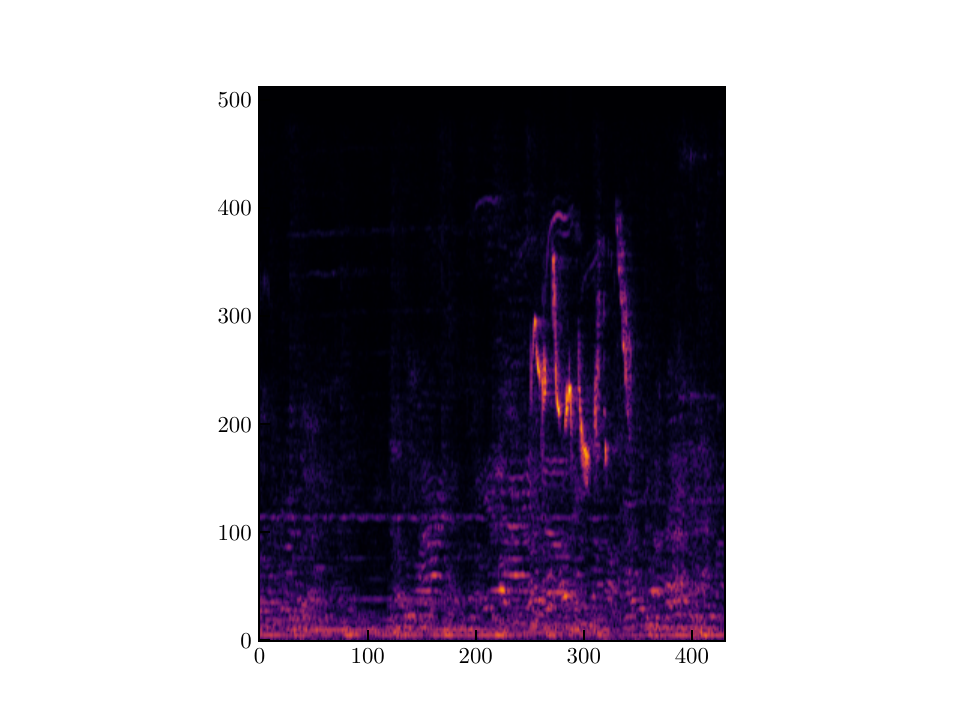} };
        \node [draw=none, fill=none, left of=ex, xshift=-1.2cm] (label)  { \rotatebox{90}{\Large GradCAM}};
        \node [draw=none, fill=none, right of=ex, xshift=4cm] (ex1)  { \includegraphics[width=0.4\textwidth]{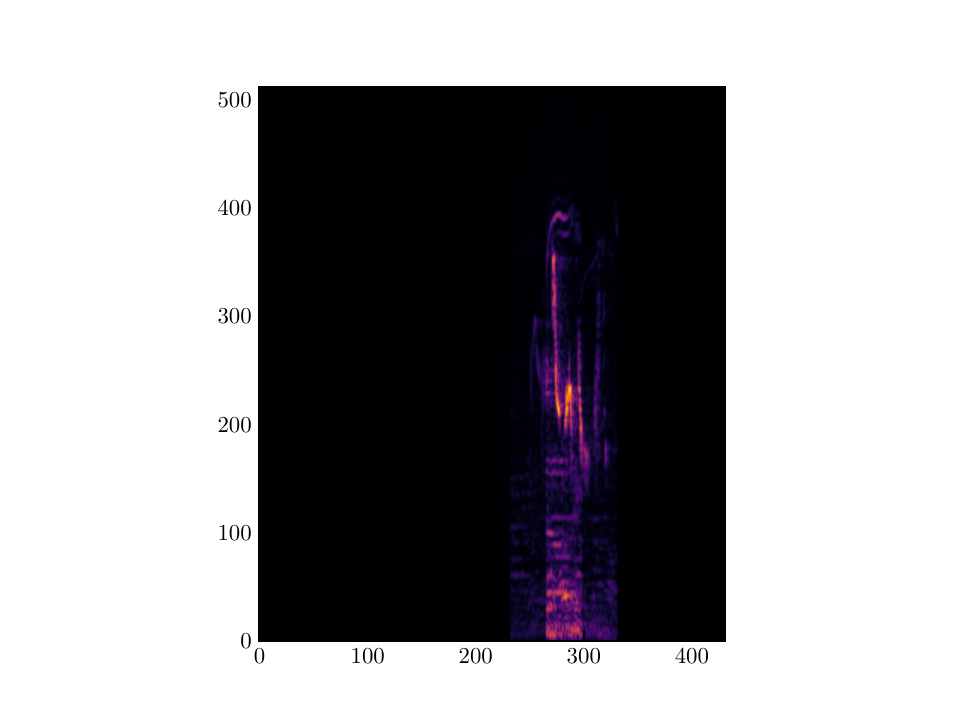} };
        \node [draw=none, fill=none, right of=ex1, xshift=4cm] (ex2)  { \includegraphics[width=0.4\textwidth]{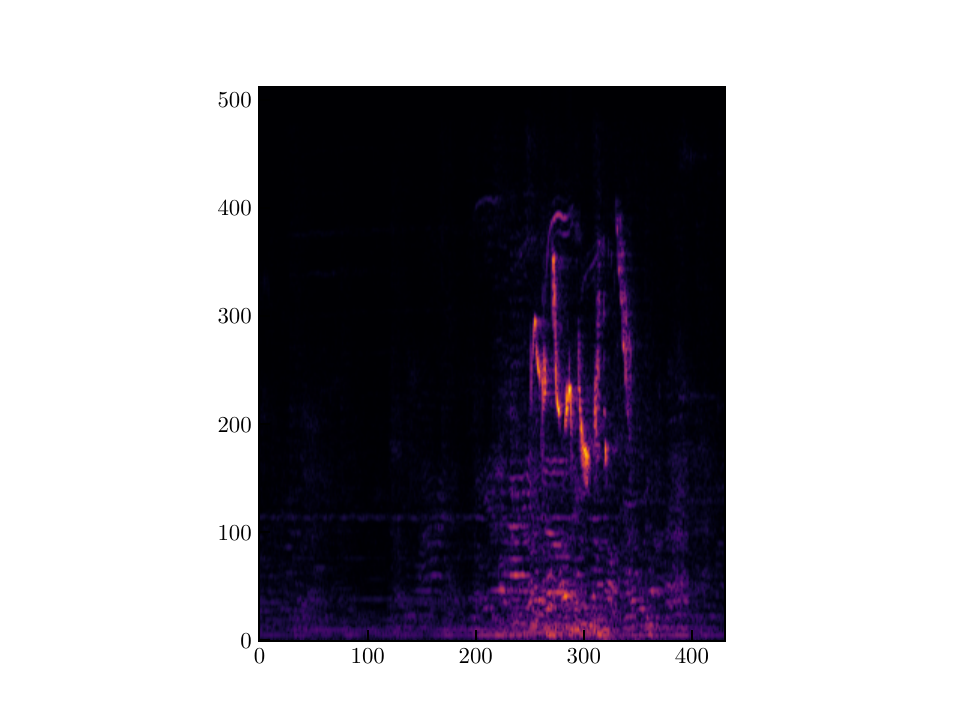} };
        \node [draw=none, fill=none, right of=ex2, xshift=4cm] (ex3)  { \includegraphics[width=0.4\textwidth]{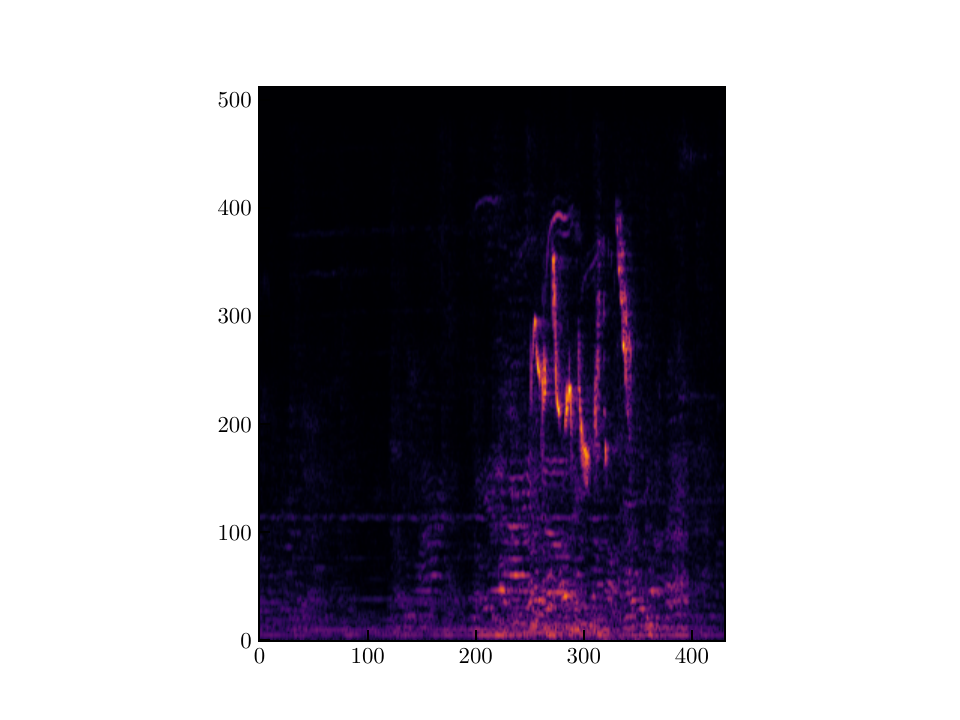} };
        \node [draw=none, fill=none, right of=ex3, xshift=4cm] (ex4)  { \includegraphics[width=0.4\textwidth]{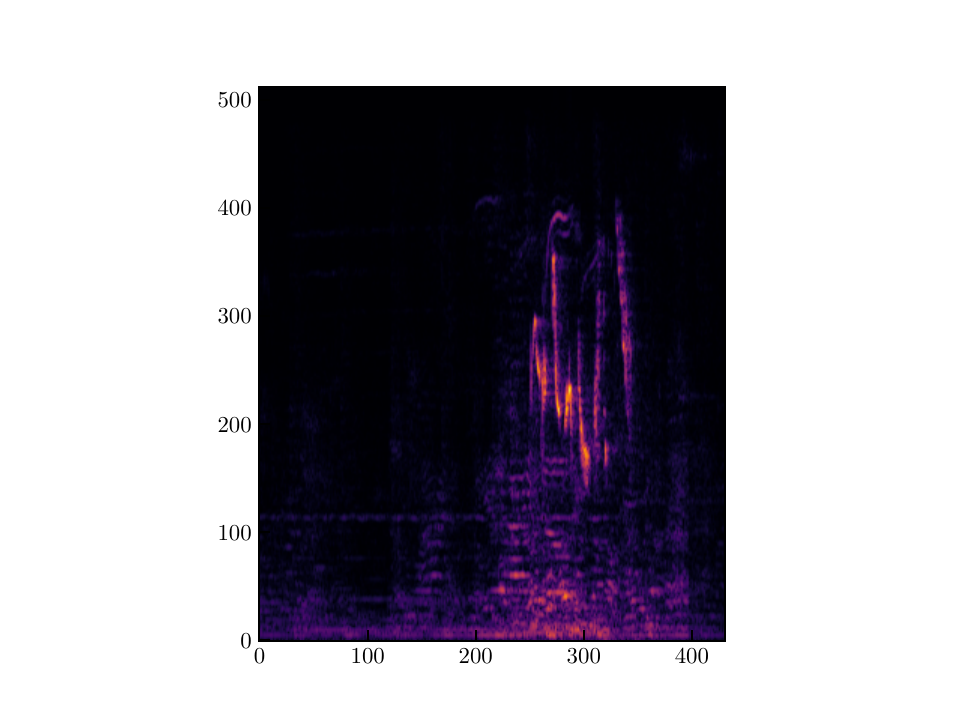} };
        \node [draw=none, fill=none, right of=ex4, xshift=4cm] (ex5)  { \includegraphics[width=0.4\textwidth]{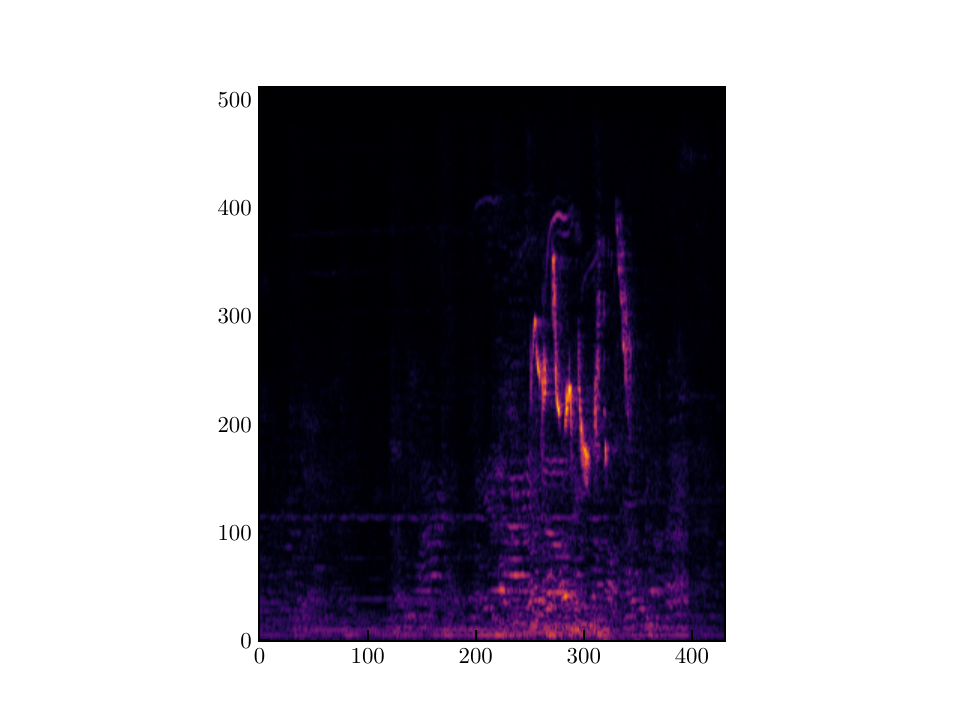} };
        \node [draw=none, fill=none, right of=ex5, xshift=4cm] (ex6)  { \includegraphics[width=0.4\textwidth]{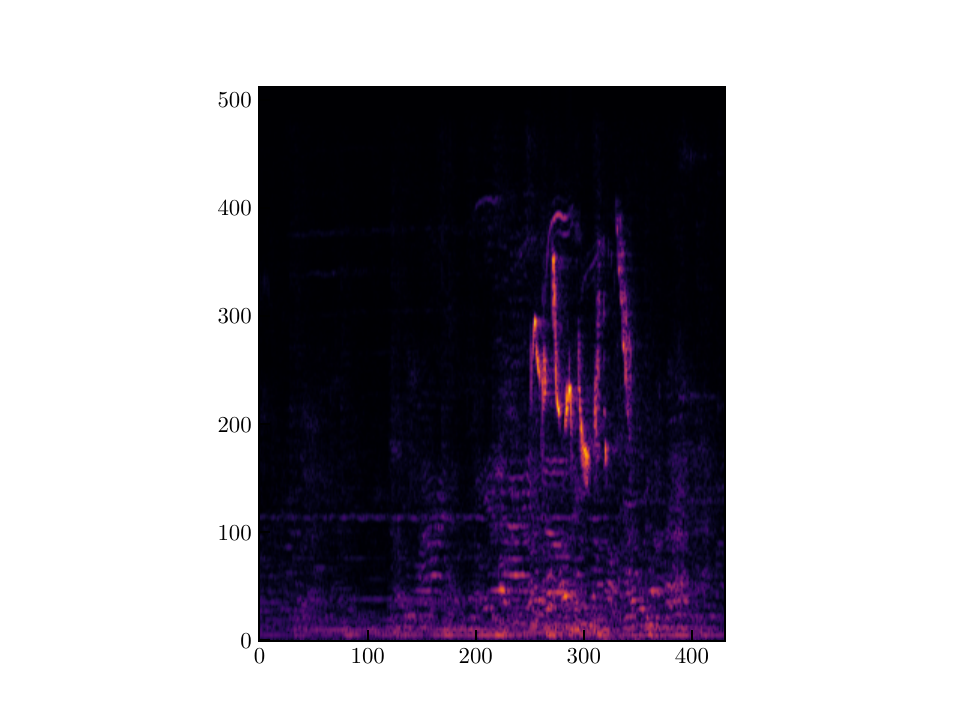} };
        \node [draw=none, fill=none, right of=ex6, xshift=4cm] (ex7)  { \includegraphics[width=0.4\textwidth]{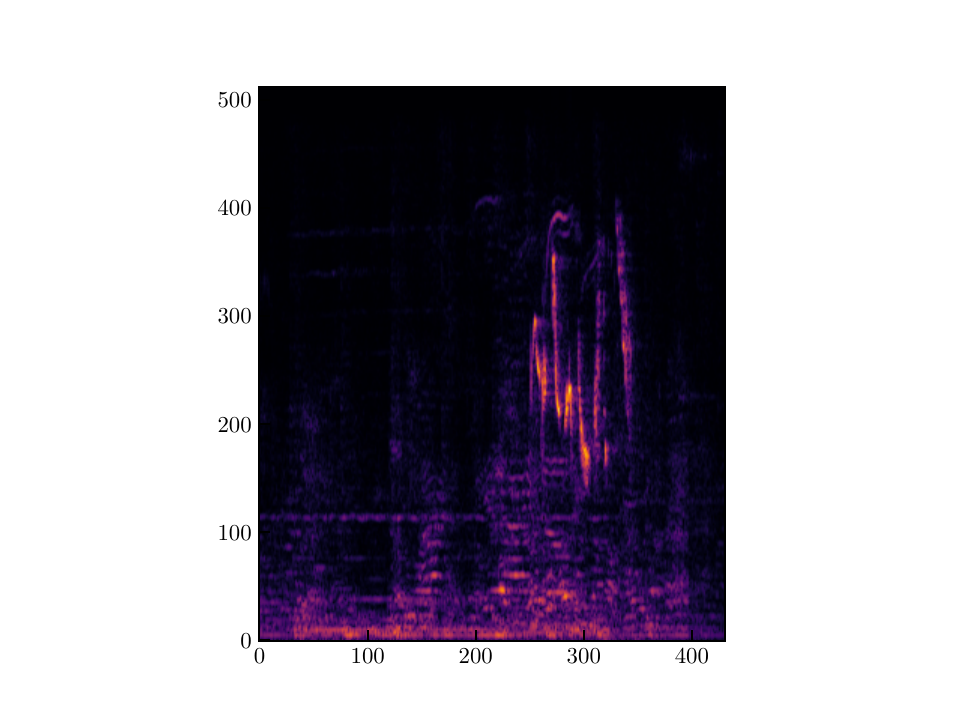} };

    \end{tikzpicture}
    }
    \vspace{-0.4cm}
    \caption{Example demonstrating the behaviour of L-MAC during the MRT test. From left to right: original sample, interpretation, and interpretations generated by randomizing the weights of the convolutional blocks starting from the logits in a cascading fashion, as suggested in \cite{adebayo2020sanity}. As expected, the interpretations are corrupted by randomizing the weights of the model. From top to bottom: L-MAC, L-MAC finetuned with $\lambda_g=4$ and CCT$=0.6$, L-MAC finetuned with $\lambda_g=4$ and $\text{CCT}=0.7$, and GradCAM.}
    \vspace{-0.1cm}
    \label{fig:cascade-viz}
\end{figure*}

In order to assess the quality of the evaluations, we have conducted a user study with 15 participants to evaluate the perceived quality of the produced interpretations. We gave the evaluators the following two instructions: 
\begin{enumerate}
    \item How well does the interpretation correspond to the part of the input audio associated with the given class? 
    \item While evaluating, please pay attention to audio quality also.
\end{enumerate}
Note that we showed the label of the predicted class to the participants. We have asked the users to rate the interpretations they listen between 0-100. We have used the open source webMushra \cite{webmushra} \href{https://github.com/audiolabs/webMUSHRA?tab=readme-ov-file}{package}.

\begin{figure*}[ht]
    \centering
    \subfloat{{ \includegraphics[width=0.43\linewidth]{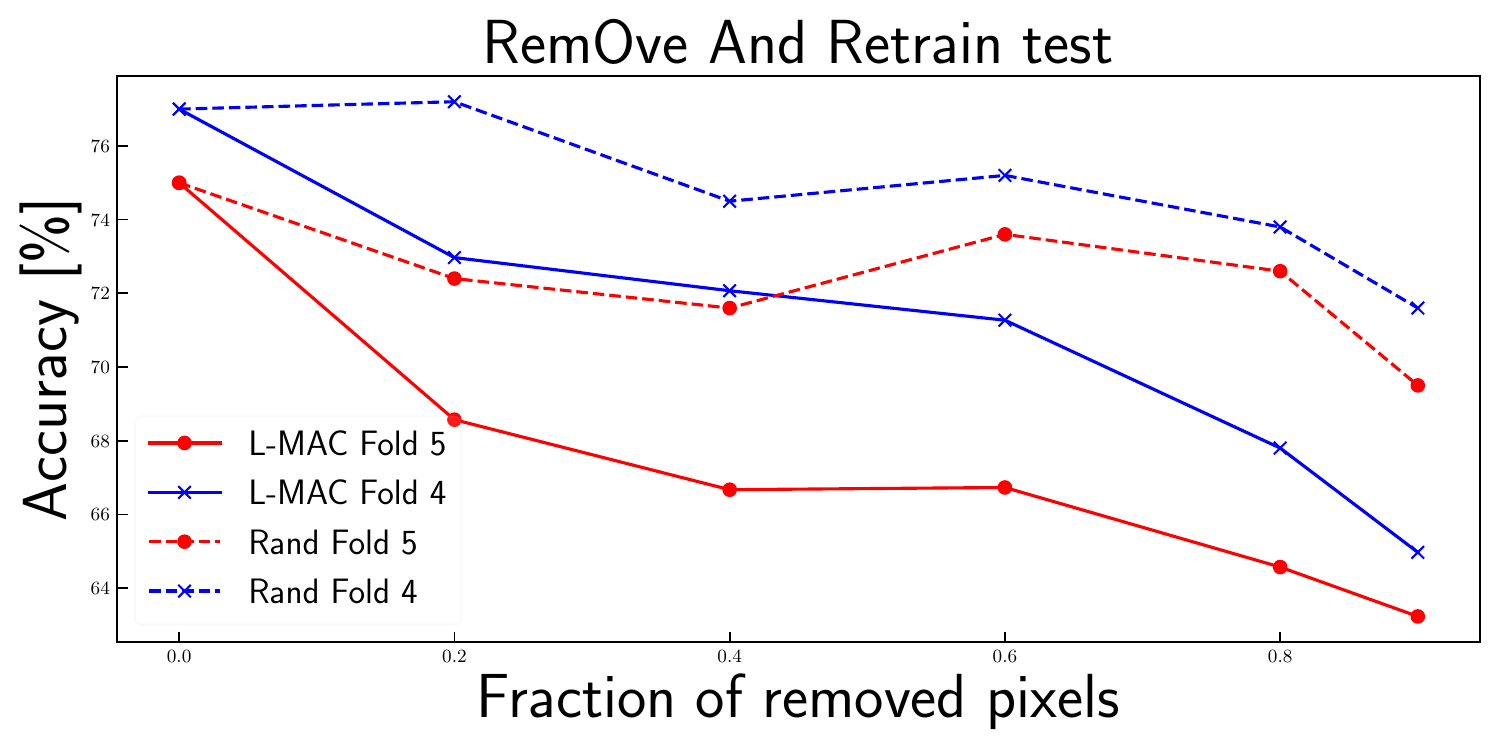} }}%
    \qquad
    \subfloat{{ \includegraphics[width=0.43\linewidth]{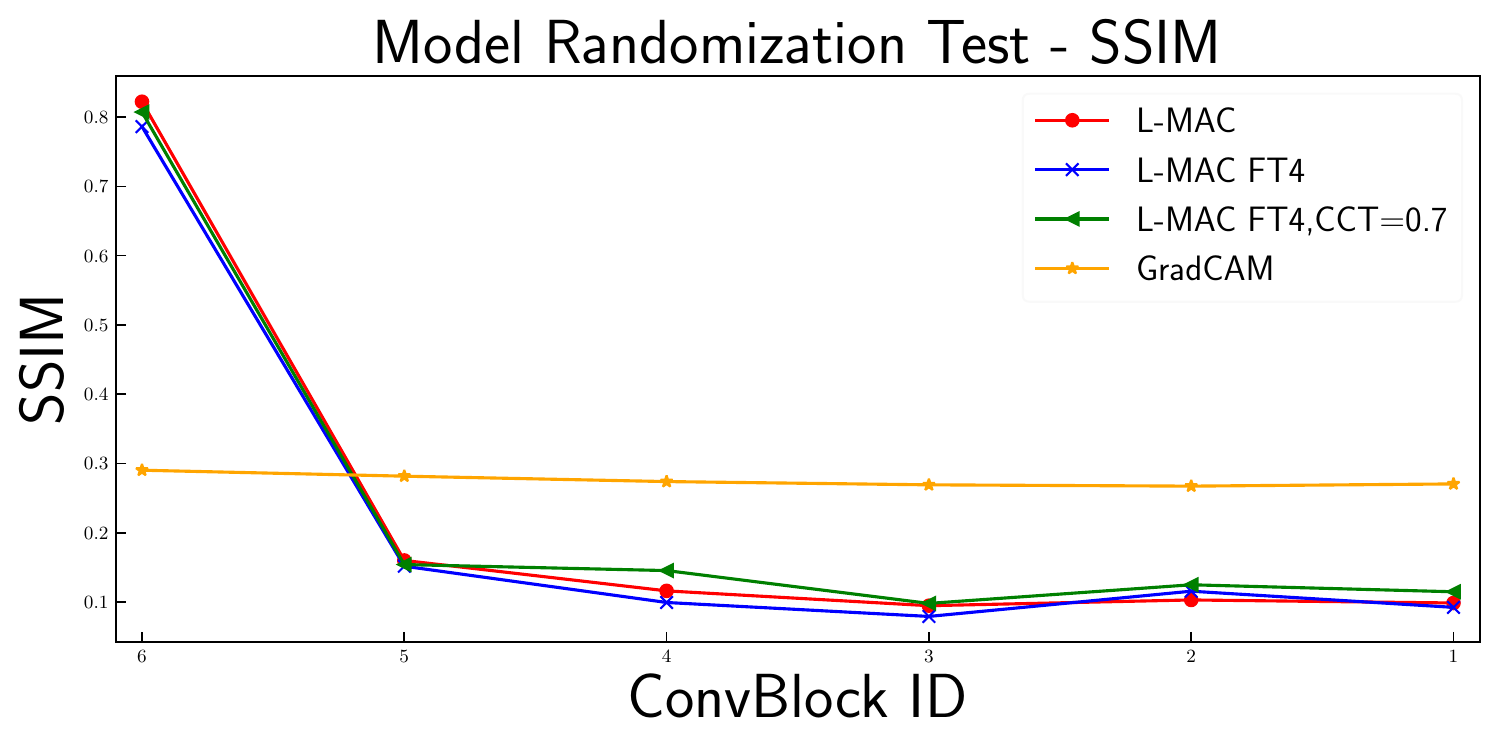} }}%
    \vspace{-0.3cm}
    \caption{Sanity checks for saliency maps: (left) RemOve And Retrain test. The presented results are the averages over three runs. The dashed line represents the random attribution baseline. (right) Structured Similarity Index (SSIM) extracted using the Model Randomization Test.}%
    \label{fig:sanitychecks}
    \vspace{-0.5cm}
\end{figure*}

In order to directly compare the perceived quality of the interpretations of L-MAC and L2I we used the audio samples provided in the L2I companion website. That is, we have downloaded the first four audio tracks and the corresponding generated interpretations from the official companion website\footnote{\url{https://jayneelparekh.github.io/listen2interpret/}} of L2I. These audio tracks are similar to the audio tracks we have used for OOD evaluation in Section \ref{sec:esc50quant}, as they are formed by mixing two audio recordings. We show the summary of this user study in Table \ref{tab:mos} in the MOS-1 column. The results indicate that, on average, users preferred the quality of the interpretations provided by L-MAC compared to the interpretations provided in the companion website of L2I. We also show the comparison of mean-opinion-scores specific to each recording in the left panel of Figure \ref{fig:esc50qualitative}. We observe that for each recording, L-MAC interpretations result in either better or comparable preference. It is worth noting that, in general, fine-tuning improves user preference compared to the standard L-MAC. 

In addition to the showcase recordings for L2I, we have also randomly picked five mixture recordings that we have created and presented these recordings to the users. In total therefore we have presented each user 9 recordings. The average preference for these recordings are shown in Table \ref{tab:mos} in the MOS-2 column. Once again, we observe that L-MAC surpasses L2I, and fine-tuning further enhances user preference. 
The results of this user study for each recording are depicted in the right panel of Figure \ref{fig:esc50qualitative}.
Some examples of interpretations can be found at our companion website\footnote{\url{https://fpaissan.github.io/lmac/}}.


\begin{table}[t]
\centering
\caption{Mean-Opinion Scores for the showcased examples from the L2I Website. L-MAC is, on average, preferred by the users.}
\label{tab:mos}
\vskip 0.15in
\begin{tabular}{lc|c}
\toprule
\textbf{Method} & MOS-1 & MOS-2 \\
\midrule
 L-MAC &  59.13 & 64.00\\
 L-MAC, FT, $\lambda_g=16$ & 59.8 & 66.7  \\ 
 L-MAC, FT, $\lambda_g=16$, CCT0.7 & \textbf{63.7} &  \textbf{67.2}  \\
 L2I & 55.1 & 50.8\\
 \bottomrule
\end{tabular}
\vspace{-0.5cm}
\end{table}

\subsection{Sanity Checks}
Beyond the quantitative and qualitative evaluation of L-MAC, we have also conducted two sanity checking experiments. First, we tested L-MAC against the Remove-and-retrain (ROAR) test proposed in \cite{hooker2019benchmark}. This test checks the classification accuracy when top-k percent of the most time-frequency bins which are deemed to be the most salient are removed, and then a classifier is retrained on the remaining portion of the input spectra. Intuitively, this sanity check verifies whether the interpreter focuses on relevant time-frequency points for the class of interest. We see in the left panel of the Figure \ref{fig:sanitychecks} that with increasing percentage in the removed portion, compared to randomly removing the masks the classification accuracy drops faster with L-MAC interpretations, which suggests that L-MAC returns interpretations on semantically relevant portion of the spectra. 


We have also conducted the cascading randomization test from \cite{adebayo2020sanity} in order to verify that L-MAC does not generate interpretations that are invariant to the classifier weights, but actually is sensitive to the randomization of the classifier weights. In Figure \ref{fig:cascade-viz}, we showcase this, where we compare the interpretations obtained when the classifier layers are randomized. We see that, as expected, with more random layers L-MAC interpretations lose their original focus. Whereas for instance with GradCAM interpretations, we observe that the interpretations basically remain insensitive to the classifier weight randomization. We furthermore quantify this finding by calculating the Structural Similarity Index (SSIM) between the interpretations generated for the original model and the interpretations generated for the model with randomized weights (as was done in \cite{adebayo2020sanity}). Similar to the visualization of the interpretations in Figure \ref{fig:cascade-viz}, we see that L-MAC interpretations quicly drop in similarity after starting from the 5'th convolutional block (5'th deepest block - note that we start randomizing from the, last layer, and then go down). Whereas, we see that the GradCAM interpretations remain practically unchanged. 



\begin{table*}[t]
\caption{Additional results obtained on the ESC50 dataset with White Noise and LJSpeech contamination. The masking is done on the STFT domain for this experiment. Note that we also indicate the mask-mean (denoted with MM)}
\label{tab:ESC50extra}
\vskip 0.15in

\centering
\resizebox{.94\textwidth}{!}{
\begin{tabular}{l|cccccccc}
\toprule
\textbf{Metric} & AI ($\uparrow$) & AD ($\downarrow$) & AG ($\uparrow$) & FF ($\uparrow$) & Fid-In ($\uparrow$) & SPS ($\uparrow$) & COMP ($\downarrow$) & MM \\
\midrule
 & \multicolumn{8}{c}{\textit{Classification on ESC50, White Noise Contamination, 38.6\% accuracy}} \\
Saliency &	0.25 &	26.31  &	0.02 &	0.05 	&0.06 & 	0.79 &	10.92 & 	0.016\\
Smoothgrad&  0.00 & 	26.37 &	0.00 & 	0.04& 	0.09& 	0.84 &	10.62 &	0.01 \\
IG &	0.75& 	25.60 & 	0.56& 	0.10 &	0.21 &	0.82 &	10.65 &	0.01 \\
L2I @ 0.2 &	0.00 &	19.41 &	0.21 &	0.11 &	0.04 	&\textbf{36.62}& 	\textbf{7.32} &	0.12 \\
GradCAM &	8.87 &	20.88 &	1.24 &	0.28 &	0.51 &	0.69 &	11.25 &	0.18\\
Guided GradCAM &	0.50 &	26.23 &	0.05 &	0.07 &	0.11 &	0.91 &	10.14 &	0.01\\
Guided Backprop &	0.25 &	26.30 	&0.02 &  0.05 &	0.07 &	0.79 &	10.92 &	0.02 \\
SHAP &	0.12 & 26.34 &	0.001 &	0.05 &	0.12 &	0.86 &	10.40 & 	0.004 \\
\textbf{L-MAC (ours)} &	\textbf{83.62} & \textbf{1.50} & \textbf{56.12} &	\textbf{0.33} &	\textbf{0.86} &	{0.92} &	{10.03} &	{0.06} \\
All-ones baseline&	0 &	0 &	0 &	0.34 &	1 &	N.A. & N.A. & 1\\ 
\midrule

 & \multicolumn{8}{c}{\textit{Classification on ESC50, LJSpeech Contamination, 79.3\% accuracy}} \\
Saliency  &0.87 &	26.00 &	0.20 &	0.06 &	0.11 &	0.75 &	11.10 &	0.02 \\
Smoothgrad 	&0.50 &	26.14& 	0.11 &	0.05 &	0.13 &	0.79 &	10.91 &	0.08 \\
IG 	&0.37 &	25.70 &	0.01 &	0.11 &	0.25 &	0.87 &	10.14 &	0.00 \\
L2I @ 0.2 & 	1.75 &	29.49 &	0.27 & 	0.15 & 	0.18 & 	0.79 & 	\textbf{9.56} & 	0.16 \\
GradCAM  &	20.37 &	13.49 &	2.63 &	0.28 &	0.73 &	0.66 &	11.33 &	0.22 \\
Guided GradCAM 	& 0.25 & 	26.10 & 	0.09 &	0.06 &	0.11 &	0.88 &	10.30 & 	0.01 \\
Guided Backprop &	0.87 & 	26.01 &	0.20 &	0.05 &	0.11 &	0.75 &	11.10 &	0.02 \\
SHAP 	&0.00 & 	26.14 &	0.00 &	0.06 &	0.16 &	0.79 &	10.81 &	0.01 \\
\textbf{L-MAC (ours)} &	\textbf{70.75} & \textbf{2.73} & \textbf{39.64} & \textbf{0.33} & \textbf{0.83} 	& \textbf{0.93} &	9.70 &	0.05 \\
All-ones baseline &  0 &	0 &	0 &	0.35 &	1 &	N/A  & N/A &	1 \\
\bottomrule
\end{tabular}
}
\end{table*}

\subsection{Additional Results on Out-of-Domain Data}
\cemcamera{In addition to the Out-of-Domain experiments conducted in Section \ref{sec:esc50quant}, we have also tested L-MAC on audio samples corrupted with white noise and speech. We have created 3dB Signal-to-noise ratio mixtures, and used samples from the LJSpeech \cite{ljspeech17} dataset for speech. In Table \ref{tab:ESC50extra}, we show these additional results on ESC50 dataset. In Appendix~\ref{app:US8k} with Table~\ref{tab:US8k}, we provide the results for this experimental setup applied on the UrbanSound8K benchmark. 
We observe that L-MAC is able to obtain better results in terms of quantitative faithfulness metrics such as AI, AD, AG, FF, and Fid-In. In this table we also report the mask-mean to indicate the size of the obtained mask (denoted with MM). We observe that even though the mask area of L-MAC is significantly smaller than GradCAM for instance, it is able to obtain considerably better metrics that measure faithfulness. Moreover, we have also calculated the scores obtained with an all-zeros mask. We observe that the FF score obtained with L-MAC is very similar to the all-ones mask which removes the entire spectrogram during  the score computation, and we also observe that the mask-mean of L-MAC is significantly smaller than 1. This indicates that the masked-out portion of the L-MAC interpretations are not very significant for the classifier.}
\vspace{-.3cm}
\section{Conclusions}

\label{sec:conclusions}
This paper introduced a novel approach, called Listenable Maps for Audio Classifiers (L-MAC), that produces post-hoc interpretations for audio classifiers. 
L-MAC employs a decoder that uses the latent representations of the black-box classifier to estimate a binary mask that effectively highlights the portions of the input audio that triggered the prediction made by the classifier.
The application of this mask to the linear-frequency-scale spectrogram enables generating interpretations that are listenable. We would like to note that our methodology is also applicable for classifiers whose input domains are not linear-frequency-scale spectra. 
We train the decoder using an objective that promotes faithfulness to the classifier decisions, that minimizes the categorical cross-entropy for masked inputs while maximizing it for masked-out inputs. 

Through an extensive experimental evaluation involving quantitative and qualitative assessments, as well as through sanity checks, our results demonstrate that L-MAC achieves significantly superior faithfulness metrics and user preference, compared to various baselines.

\vspace{-.3cm}
\section*{Impact statement}
\cemcamera{
This paper proposes a methodology to improve neural network interpretability. We do not foresee a direct, socially negative impact. On the contrary, the proposed methodology may facilitate socially beneficial applications of machine learning-based audio processing by enhancing the human trust in neural network decisions. 
}
\vspace{-.2cm}
\section*{Acknowledgements}
\cemcamera{We thank Geraldin Nanfack and Eugene Belilovsky for insightful discussions. This research was enabled in part by support provided by Calcul Québec and the Digital Research Alliance of Canada. }


\bibliography{example_paper}
\bibliographystyle{icml2024}

\newpage
\appendix
\onecolumn

\section{Decoder structure illustration}
\label{app:diag}
\begin{figure}[h]
    \centering
    \includegraphics[width=\textwidth]{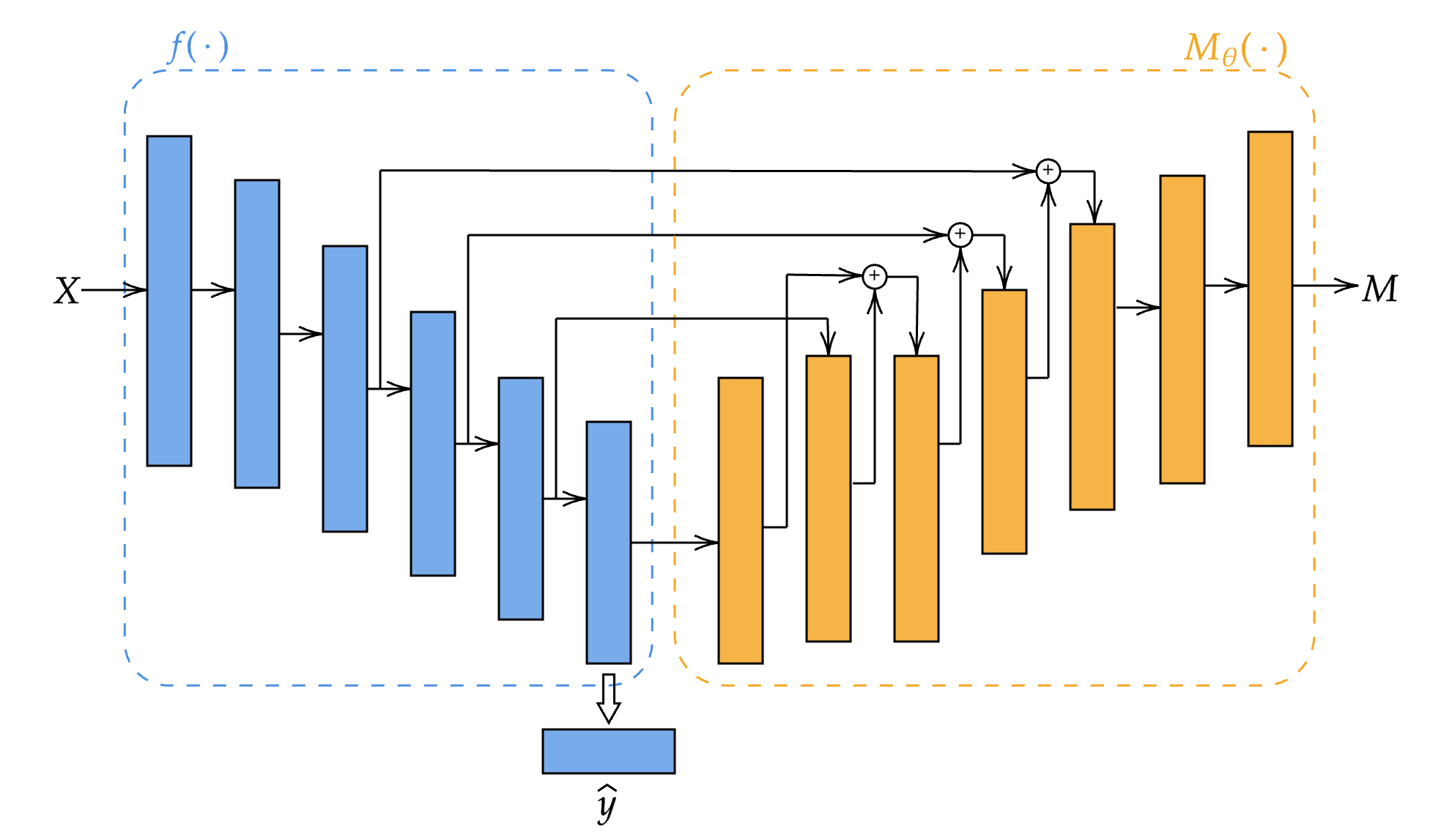}
    \caption{Diagram of the decoder neural network $M_\theta(\cdot)$. The representations from the classifier $f(\cdot)$ are fed through different layers of the decoder $M_\theta(\cdot)$ via skip connections.}
    \label{fig:unet_diag}
\end{figure}

In Figure~\ref{fig:unet_diag} we show the architecture of the decoder $M_\theta(\cdot)$ employed in our experiments.

\section{Results on UrbanSound8k Dataset}
\label{app:US8k}

\cemcamera{We provide additional results for the In-Domain and Out-of-Domain experiments on the UrbanSound8k dataset \cite{us8k}, in addition to the experiments conducted on the ESC50 dataset in the mainbody of the paper. In Table \ref{tab:US8k}, we first report the results obtained on the In-Domain experiments, and then for the Out-of-Domain experiments where we contaminate the samples with other samples from the US8k dataset, White Noise, and speech (the same experimental setup as the setup on the ESC50). We observe that on both the ID experiment and on different OOD experiments, L-MAC is able to outperform or obtain very similar results compared to the baselines. We also observe that L-MAC masks are in general smaller compared to the ones obtained by comparably performing models such as GradCAM or L2I, which suggests that L-MAC is able to obtain masks that are not trivial.} 

\begin{table*}[t]
\caption{Additional experiments on UrbanSound8k Dataset}
\label{tab:US8k}
\vskip 0.15in

\centering
\resizebox{.94\textwidth}{!}{
\begin{tabular}{l|cccccccc}
\toprule
\textbf{Metric} & AI ($\uparrow$) & AD ($\downarrow$) & AG ($\uparrow$) & FF ($\uparrow$) & Fid-In ($\uparrow$) & SPS ($\uparrow$) & COMP ($\downarrow$) & MM \\
\midrule
 & \multicolumn{8}{c}{\textit{Classification on US8k, WHAM! Noise Contamination (ID experiment), 82.8\% accuracy}} \\
Saliency &	0.40 &	23.86 &	0.26 & 	0.072 &	0.12 &	0.30 & 	5.69 & 	0.02 \\
Smoothgrad &	0.00 &	26.86 &	0.00 &	0.04 &	0.04 & 	0.30 & 	5.69 & 0.01 \\
IG 	& 0.25 & 	23.70 & 	0.15 &	0.10 & 	0.12 & 	\textbf{0.40} & 	\textbf{5.24} & 	0.005 \\
L2I @ 0.2 &	0.55 & 	19.59 &	0.28 & 	0.17 & 	0.18 & 	0.24 & 	5.67 & 	0.21 \\
GradCAM  &7.68 & 	6.94 & 	4.04 & 	0.31 & 	0.39 & 	0.13 & 	5.96 & 	0.29 \\
Guided GradCAM &	0.40 &	24.29 &	0.26 & 	0.06 & 0.10 & 0.33 & 	5.58 & 	0.01 \\
Guided Backprop &	0.40 & 	23.86 &	0.26 &0.07 &0.12 & 	0.30 & 	5.69 & 	0.02 \\
SHAP &	0.10 &	26.32 &	0.03 & 	0.05 & 	0.06 & 	0.33 & 	5.59 & 	0.01 \\
\textbf{L-MAC (ours)} &	\textbf{19.09} & \textbf{2.51} & \textbf{13.57} & \textbf{0.40} & \textbf{0.46} & 0.32 &	5.55 &0.14 \\
All-ones baseline 	& 0 &	0 &	0 &	0.42 &	1 &	N/A &	N/A &	1 \\
\midrule
 & \multicolumn{8}{c}{\textit{Classification on US8k, US8k Contamination, 82.6\% accuracy}} \\

Saliency &	0.40 &	48.24 & 	0.23 & 	0.10 & 	0.24 & 	0.61 & 	11.35 & 	0.056 \\
Smoothgrad &0.00 &	54.65 & 	0.00 &	0.06 & 	0.09 & 	0.60 & 	11.40 & 	0.023 \\
IG &	0.51 &	46.31 & 	0.20 & 	0.15 & 	0.30 & 	\textbf{0.80} & 	\textbf{10.39} & 	0.01 \\
L2I @ 20 &	2.88 & 	27.84 &	2.01 & 	0.27 & 	0.38 & 	0.50 & 	11.05 & 	0.37 \\
GradCAM  & 14.65 & 	9.71 & 	6.42 & 	0.61 & 	0.86 & 	0.18 & 	12.02 & 	0.60 \\
Guided GradCAM 	& 0.40 & 49.61 & 0.19 & 0.10 & 	0.21 & 	0.66 & 	11.17 & 	0.03 \\
Guided Backprop  &0.40 & 	48.24 & 0.23 & 	0.10 & 	0.24 & 	0.61 & 	11.35 & 0.04 \\
SHAP  & 0.05 & 	53.66 &	0.01 &  0.07 & 	0.12 & 	0.67 & 	11.17 & 	0.01 \\
\textbf{L-MAC (ours)} & \textbf{27.12} & \textbf{9.58} & \textbf{20.04} &	\textbf{0.77} & \textbf{0.85} & 0.71 & 	10.75 & 0.24 \\
All-ones baseline & 0 &	0 &	0 	&0.86 &	1 &	N/A &	N/A &	1 \\
\midrule
 & \multicolumn{8}{c}{\textit{Classification on US8k, White Noise Contamination, 48.5\% accuracy}} \\
Saliency & 0.56 & 48.42 & 0.16 & 0.13 & 0.14 & 0.57 & 11.50 & 0.04 \\
Smoothgrad & 0.00 & 48.96 & 0.00 & 0.07 & 0.10 & 0.66 & 11.24 & 0.02 \\
IG & 0.85 & 47.92 & 0.30 & 0.15 & 0.16 & 0.63 & 11.21 & 0.02 \\
L2I @ 0.2 & 27.45 & 32.73 & 23.45 & 0.25 & 0.42 & 0.32 & 11.81 & 0.41 \\
GradCAM & \textbf{37.37} & 16.69 & \textbf{31.29} & 0.45 & 0.69 & 0.24 & 11.95 & 0.61 \\
Guided GradCAM & 0.51 & 48.26 & 0.27 & 0.12 & 0.14 & 0.62 & 11.31 & 0.03 \\
Guided Backprop & 0.56 & 48.41 & 0.16 & 0.13 & 0.13 & 0.59 & 11.44 & 0.04 \\
SHAP & 0.15 & 48.51 & 0.02 & 0.09 & 0.12 & \textbf{0.72} & \textbf{10.98} & 0.01 \\
\textbf{L-MAC (ours)} & 31.62 & \textbf{16.31} & 22.4 & \textbf{0.81} & \textbf{0.71} & 0.39 & 11.72 & 0.39 \\
All-ones baseline & 0 & 0 & 0 & 0.74 & 1 & N/A & N/A & 0 \\

 \midrule
  & \multicolumn{8}{c}{\textit{Classification on US8k, LJSpeech Contamination, 88.7\% accuracy}} \\
Saliency & 0.30 & 50.60 & 0.11 & 0.08 & 0.25 & 0.58 & 11.44 & 0.04 \\
Smoothgrad & 0.00 & 52.68 & 0.00 & 0.05 & 0.19 & 0.61 & 11.39 & 0.02 \\
IG & 0.35 & 50.13 & 0.18 & 0.10 & 0.26 & \textbf{0.79} & \textbf{10.43} & 0.01 \\
L2I @ 0.20 & 1.21 & 41.91 & 0.72 & 0.23 & 0.27 & 0.55 & 10.50 & 0.23 \\
GradCAM & 13.29 & \textbf{7.16} & 6.20 & 0.53 & \textbf{0.91} & 0.25 & 11.93 & 0.61 \\
Guided GradCAM & 0.30 & 51.18 & 0.14 & 0.08 & 0.23 & 0.64 & 11.27 & 0.03 \\
Guided Backprop & 0.30 & 50.6 & 0.11 & 0.09 & 0.25 & 0.58 & 11.44 & 0.04 \\
SHAP & 0.05 & 52.15 & 0.01 & 0.06 & 0.18 & 0.67 & 11.16 & 0.01 \\
\textbf{L-MAC (ours)} & \textbf{18.18} & 9.91 & \textbf{11.28} & \textbf{0.90} & 0.86 & 0.69 & 10.91 & 0.26 \\
All-ones baseline & 0 & 0 & 0 & 0.83 & 1 & N/A & N/A & 0 \\

\bottomrule
\end{tabular}
}
\end{table*}

\end{document}